\documentclass[useAMS,usenatbib]{mn2e}
\input{epsf}
\usepackage{amssymb}
\usepackage{amsmath}
\usepackage{natbib}
\usepackage{color}
\usepackage{journals}
\usepackage{graphicx}
\usepackage{caption}
\usepackage{subcaption}

\newbox\grsign \setbox\grsign=\hbox{$>$} \newdimen\grdimen \grdimen=\ht\grsign
\newbox\simlessbox \newbox\simgreatbox
\setbox\simgreatbox=\hbox{\raise.5ex\hbox{$>$}\llap
     {\lower.5ex\hbox{$\sim$}}}\ht1=\grdimen\dp1=0pt
\setbox\simlessbox=\hbox{\raise.5ex\hbox{$<$}\llap 
     {\lower.5ex\hbox{$\sim$}}}\ht2=\grdimen\dp2=0pt

\newcommand{\hMpc}{{\ifmmode{h^{-1}{\rm Mpc}}\else{$h^{-1}$Mpc }\fi}}
\newcommand{\hGpc}{{\ifmmode{h^{-1}{\rm Gpc}}\else{$h^{-1}$Gpc }\fi}}
\newcommand{\hkpc}{{\ifmmode{h^{-1}{\rm kpc}}\else{$h^{-1}$kpc }\fi}}
\newcommand{\hMsun}{{\ifmmode{h^{-1}{\rm {M_{\odot}}}}\else{$h^{-1}{\rm{M_{\odot}}}$}\fi}}
\newcommand{\Msun}{{\ifmmode{{\rm {M_{\odot}}}}\else{${\rm{M_{\odot}}}$}\fi}}

\title[Consistent extinction]{Consistent extinction model for type Ia supernovae in Cepheid-based calibration galaxies and its impact on $H_{0}$}
\author[R. Wojtak]{Rados{\l}aw~Wojtak$^{1}$\thanks{E-mail: radek.wojtak@nbi.ku.dk}, Jens Hjorth$^{1}$\\
$^{1}$DARK, Niels Bohr Institute, University of Copenhagen, Jagtvej 155, 2200 Copenhagen, Denmark  \\
}
\begin{document}

\maketitle

\begin{abstract}

The most recent SH0ES measurement of the Hubble constant employs corrections of type Ia supernova magnitudes due to extinction in their host galaxies. These corrections are estimated using a probabilistic model which is trained on Hubble flow ($z>0.03$) supernovae and extrapolated to the calibration galaxies (those with observed Cepheids), despite the fact that the latter are selected based on criteria favouring disky and dust-rich systems. We show that this standard approach underestimates the brightness of reddened supernovae in the high stellar-mass ($M_{\star}>10^{10}M_{\odot}$) calibration galaxies. This can be traced back to the fact that for these galaxies, a low total-to-selective extinction coefficient ($R_{\rm B}\sim 3$) is assumed, while for the low stellar-mass analogues a more standard $R_{\rm B}\sim 4$ is adopted. We propose a minimalistic modification of the extinction model in the calibration galaxies in order to alleviate this systematic effect. The modification is twofold and it involves: (i) the same, Milky Way-like distribution of $R_{\rm B}$ (with mean $R_{\rm B}$ of $4.3$ -- consistent with the extinction curve used for colour corrections of the Cepheids --  and scatter $0.4$) and (ii) a modified shape of the $E(B-V)$ reddening distribution while keeping the same effective slope of the supernova peak magnitude-colour relation and the same mean $E(B-V)$ reddening as measured for supernovae in the Hubble flow. We show that this new approach yields a significantly better fit ($\Delta BIC=-11$) to the calibration data and results in a lower value of $H_{0}$. Our result is $H_{0}=70.5\pm1$~km~s$^{-1}$~Mpc$^{-1}$ implying a reduction of the Hubble constant tension from $5.2\sigma$ to $2.8\sigma$.
\end{abstract}


\section{Introduction}\label{introduction}

The Hubble constant ($H_{0}$) measured in the Supernovae and H0 for the Dark Energy Equation of State (SH0ES) program 
using observations of Cepheids and type Ia supernovae \citep{Riess2022} is 
$5.6\pm{1.15}$~km~s$^{-1}$~Mpc$^{-1}$ higher than the value derived from the Planck observations of the cosmic microwave background (CMB) 
radiation assuming a standard flat $\Lambda$CDM cosmological model \citep{Planck2020_cosmo}. The discrepancy between the two determinations 
of the present expansion rate has reached a $5\sigma$ significance. It is hypothesised that the tension may be an observational signature of 
new physics beyond the standard cosmological framework. Consequently, tremendous effort has been put into exploring a wide range of possible modifications to the standard cosmological model, 
which could resolve the problem \citep[][]{DiVal2021}. However, it appears increasingly unlikely to find a compelling solution to the $H_{0}$ tension 
when one considers a whole cosmology view supported by all relevant observational constraints \citep{Linder2023}. 
Extra caution should also be taken when new models appear to be favoured based solely on evaluating a difference 
between best fit values of a single parameter derived from different data sets \citep{Cor2023}. 
It is now widely accepted that observations of baryon acoustic oscillations (BAO), which effectively gauge distance scales based on pre-recombination physics, 
hinder attempts to resolve the Hubble constant tension by means of altering the expansion history at $z\lesssim 2$ 
\citep[the so-called late-time solutions;][]{Schoneberg2022,Arendse2020,Pogosian2022}. The only way to increase the CMB-inferred $H_{0}$ and remain 
consistent with the BAO observations at the same time is to decrease the sound horizon by means of modifying the standard cosmological model before recombination. 
The most promising class of solutions along this path involves early dark energy \citep[see e.g.][]{Poulin2019,Niedermann2020}. The main drawback of this approach is that the current CMB observations 
alone do not seem to favour early dark energy as an extension to the standard cosmological model \citep[]{Poulin2023,McDon2023,Efst2023,Vagnozzii2021}. 
Furthermore, cosmological models imposing early dark energy tend to give rise to discrepancies with other cosmological data sets \citep[][but see also \citealt{Smith2021,Niedermann2021}]{Hill2020,Ivanov2020,Vagnozzi2023,Goldstein2023,Jedamzik2021}.

Apparent differences between Hubble constant determinations based on alternative distance calibrations to Cepheids, e.g. the Tip of Red Giant Branch \citep[TRGB;][]{Freed2019,Freedman2021}, surface brightness fluctuations \citep[SBF;][]{Khetan2021,Jensen2021}, or different 
choices of filters to measure supernova light curve parameters \citep{Uddin2023} may be tentatively interpreted as an indication 
of unaccounted for systematic errors. Perhaps the most significant and persistent discrepancy lies between the TRGB-based 
measurement obtained by \citet[][]{Freed2019} and the Cepheid-based determination of \citet{Riess2022}. The ongoing effort to test possible systematic effects behind 
these two $H_{0}$ measurements is primarily focused on the methods, models and observations used to propagate direct geometric distance estimates to nearby galaxies 
hosting type Ia supernovae \citep{Moertsell2022a,Moertsell2022,Freedman2023,Majaess2024}, with an important role of the James Webb Space Telescope \citep[JWST;][]{Freedmann2023a,Riess2023}.  
A wide range of potential systematic effects related to the general framework of supernova analysis -- i.e.\ photometric calibrations, light curve fitting, the Milky Way extinction and the accuracy of redshifts (including peculiar velocity corrections) -- have been shown to have negligible impact on the Hubble constant determination \citep[][but see also \citealt{Steinhardt2020}]{Carr2022,Brout2022,Petersen2022}. However, potential biases arising from differences between supernova environments in the calibration galaxies and the Hubble flow have been largely neglected.

The accuracy of the distance ladder technique with type Ia supernovae implicitly relies on the assumption that the population properties of supernovae and 
their local environments in the calibration galaxies and the Hubble flow match with adequate precision. However, a clear observational signature 
that this condition is not fully met in the case of the Cepheid-based distance calibration was found 
by \citet{Wojtak2022}, based on an analysis of the SuperCal supernova compilation \citep{Scolnic2015} and Cepheid data from \citet{Riess2016}. 
This supernova data set was used in the SH0ES measurement of the Hubble constant of \citet[][and its updates following 
subsequent improvements of geometric distance estimates]{Riess2016} resulting in a $4.2\sigma$ significance of the Hubble constant tension 
\citep{Riess2019,Riess2021}. \citet{Wojtak2022} showed that the supernovae in the calibration galaxies \citep[19 galaxies from][]{Riess2016} 
require a substantially steeper colour correction than supernovae in the Hubble flow, with a (2.2--3.8)$\sigma$ significance depending on the treatment 
of the intrinsic scatter in supernova peak magnitudes. The most straightforward interpretation of this result is that the calibration galaxies exhibit a stronger extinction 
than supernova hosts in the Hubble flow. A simple way to minimise the related systematic errors in the $H_{0}$ determination is to (i) fit supernova peak magnitude--colour relations 
independently in the calibration sample and the Hubble flow and (ii) propagate distance measurements using standardised peak magnitudes linked 
to a colour which closely matches supernova intrinsic colours (minimum dust reddening). Using recent estimates of supernova intrinsic colours from Bayesian 
hierarchical modelling \citep[see e.g.][]{Popovic2021,Wojtak2023} the best fit Hubble constant decreases to $H_{0}\approx 70$~km~s$^{-1}$~Mpc$^{-1}$ \citep[see Figure 4 of ][]{Wojtak2022} 
reducing the Hubble constant tension by a factor of 2.

The main goal of the present paper is to test if the observational evidence for higher extinction in the calibration galaxies demonstrated 
by \citet{Wojtak2022} holds for the new Cepheid observations from \citet{Riess2022} and the associated Pantheon+ supernova 
compilation \citep{Brout2022}. The new data are different in several respects. The new calibration sample is nearly twice as large as the \citet{Riess2016} 
sample and the old Cepheid data have been reprocessed. For the supernova data, the main changes include improved 
light curve fits \citep{Scolnic2022_lc}, photometric calibration \citep{Brout2022_fra} and redshifts \citep{Carr2022}, as well as new supernovae observed recently both in some calibration galaxies and in the Hubble flow. The most crucial addition to the Pantheon+ compilation, however, is a probabilistic model of extinction in supernova host galaxies, implemented through a colour-dependent bias and intrinsic scatter model. The dust model developed for the bias correction was obtained from forward modelling of observed supernova colours and Hubble residuals at $z>0.03$ \citep[][herafter P23]{Popovic2021}. It was applied to the calibration 
sample assuming that the population properties of dust reddening and extinction in the calibration sample and the Hubble flow 
match sufficiently well given the current precision of the Hubble constant measurement. The aim of this study is to test whether this implicit 
extrapolation, i.e. applying the model to a subclass of supernova host galaxies from the training data, 
can be substantiated by the Cepheid data or whether it leads to a similar signature of underestimated extinction as that shown 
by \citet{Wojtak2022}. We emphasize that any potential difference between the calibration and Hubble flow galaxy samples in terms of extinction can arise from the fact that the former consists solely of late-type galaxies and thus it is not a random representation of the latter.

The P23 dust model employed in the Pantheon+ compilation provides a physically motivated interpretation of observed supernova 
Hubble residuals and their dependence on supernova colour \citep{Brout2021,Duarte2023,Rose2022}. It ascribes two different extinction properties 
to supernova host galaxies with high ($M_{\star}>10^{10}M_{\odot}$) or low stellar masses ($M_{\star}<10^{10}M_{\odot}$), 
with an average total-to-selective coefficient in the $B$-band of $R_{\rm B}\approx 3$ for the former and  $R_{\rm B}\approx 4$ for the latter \citep{Popovic2021}. 
The average difference between Hubble residuals in different stellar mass bins is well known and was 
previously included 
as the so-called mass-step correction in type Ia supernova standardisation \citep[see e.g.][]{Kelly2010,Smith2020}.

The colour-dependent mass-step underlying the P23 extinction model may actually be an emergent phenomenon resulting from 
different proportions of young (dust-rich) and old (dust-free) stellar populations in high and low stellar-mass galaxies \citep[see e.g.][]{Kauffmann2003}. 
Recent two-population hierarchical Bayesian modelling, in which extinction properties of young (old) stellar environments are indirectly probed through 
observationally associated type Ia supernovae with slowly (fast) declining light curves \citep{Rigault2013}, seems to corroborate this scenario \citep{Wojtak2023}. 
Here, one should also expect the mass-step correction to be partially driven by differences in supernova intrinsic properties, for which 
observational signatures were recently demonstrated by \citet[][]{Grayling2024} and 
\citet[][]{Duarte2023}.
While an understanding of the mass-step correction is essential as a basis for a sound theoretical framework for the cosmic distance ladder with type Ia supernovae, a complete physical interpretation is yet to be established. However, the current 
models such as P23 can effectively be used as an accurate description of supernova Hubble residuals as long as the population properties of 
supernovae and host galaxies closely match those of the training data set. In this work, we assume that the P23 model by construction provides an 
accurate method to standardise type Ia supernovae in the Hubble flow, but we test its applicability to the calibration galaxies.

Possible differences between the calibration galaxies and galaxies in the Hubble flow in terms of dust reddening and extinction may occur due to the fact 
that the former are biased: luminous Cepheids are found in star-forming and thus dusty galaxies. \citet{Riess2022} attempted to mitigate 
this bias by selecting similar late-type galaxies in the Hubble flow, based on visual assessment of morphological types from the best optical images, and using the resulting restricted supernova sample for the $H_{0}$ determination. The adopted selection was shown to have a negligible impact on the estimation of the Hubble constant \citep{Riess2022}. This implies two possible scenarios: either 
the dust reddening and extinction properties are nearly the same in each of the three galaxy samples (the calibration galaxies, the restricted Hubble flow and 
the full Hubble flow) or the employed host galaxy selection in the Hubble flow is not precise enough to capture the exact properties of dust reddening 
and extinction in the calibration galaxies. In this study, we explore the latter possibility by means of testing directly alternative extinction models against the calibration data.

The outline of the paper is a follows. In section 2 we describe the supernova and Cepheid data, and the P23 dust model implemented through the bias corrections and 
covariance matrix of type Ia supernovae in the Pantheon+ compilation. In section 3 we carry out consistency tests of the dust model using the calibration data. 
In section 4 we develop a new dust model, test it against the calibration data and derive a new estimate of the Hubble constant. Discussion and summary 
follow in sections 4 and 5.

\section{Pantheon+ data and the dust model}

We use measurements of type Ia supernova corrected peak magnitudes from \cite{Brout2022} 
and distance moduli of 37 calibration galaxies obtained from SH0ES observations of Cepheids \citep[][]{Riess2022}. 
Our analysis focuses on 42 distinct supernovae (77 including duplicates) in the calibration sample. 
We use the publicly available covariance matrix which includes both statistical and systematic uncertainties 
of supernova corrected magnitudes and distances moduli from Cepheids\footnote{https://github.com/PantheonPlusSH0ES}.

\subsection{Supernova corrected magnitudes}

Light curves of type Ia supernovae from the Pantheon+ catalogue are fitted using the SALT2 model \citep{Guy2007}, 
with retrained model parameters from \citet{Taylor2021}. In this approach, every normal type Ia supernova is described by three 
light curve parameters: the apparent rest-frame $B$-band peak magnitude $m_{\rm B}$, the dimensionless stretch parameter $x_{1}$ 
quantifying the width of the light curve and the colour parameter $c$. Distance moduli are inferred from the light curve parameters 
assuming a model of supernova standardisation given by
\begin{equation}
\mu\equiv m_{\rm B,corr}-M_{\rm B}=m_{\rm B}-M_{\rm B}+\alpha x_{1}-\beta c -\delta(c,M_{\star}),
\label{mb_main}
\end{equation}
where $M_{\rm B}$ is the rest-frame $B$-band absolute magnitude, $m_{\rm B, corr}$ is the corrected peak magnitude and $\delta(c,M_{\star})$ 
is a bias correction, which is a function of supernova colour and host galaxy stellar mass \citep{Brout2022}. The 
two linear functions of $x_{1}$ and $c$ were proposed by \citet{Tripp1998} as an empirical model which reduces the bulk of observational 
scatter on supernova Hubble diagrams. The model coefficients are determined directly from observations. The term $\delta(c,M_{\star})$ 
corrects for three effects which cannot be accounted for by the Tripp model. Firstly, it includes second-order corrections 
arising from full forward modelling of supernova colour parameters and observed magnitudes based on physically motivated models of 
dust and supernova intrinsic colours \citep{Popovic2021}. We outline all relevant details of that model in the following section. Secondly, the bias term also
accounts for observational selection effects estimated by simulating the actual survey strategies with the SuperNova Analysis (SNANA) program \citep[][]{Kessler2009}. Finally, $\delta$ can also include the so-called mass-step correction which assigns 
different absolute absolute magnitudes for supernovae in host galaxies with stellar masses higher or lower than $10^{10}M_{\odot}$. 
The difference between the absolute magnitudes is measured from observations and 
is in a range between 0.05 and 0.08 mag \citep[see e.g.][]{Smith2020,Scolnic2018}. In the Pantheon+ compilation, however, the mass-step 
is incorporated directly in the dust model which accounts for this effect by postulating different extinction properties in the two stellar mass bins of the mass-step. 
Lower total-to-selective extinction with $R_{\rm B}\approx 3$ is ascribed to brighter supernovae in the high stellar-mass ($M_{\star}>10^{10}M_{\odot}$) 
host galaxies, whereas higher extinction with $R_{\rm B}\approx 4$ to fainter supernovae in the low stellar-mass host galaxies \citep{Brout2021,Popovic2021}.

Figure~\ref{HD_ori_b} shows the bias term $\delta$ from the Pantheon+ catalogue, as a function of colour parameter $c$. It is 
apparent that supernovae in the calibration sample or the Hubble flow closely follow two distinct branches of $\delta(c)$. 
These two branches result from modelling dust reddening and extinction in two separate bins of supernova host galaxy stellar masses \citep{Brout2021,Popovic2021}. 
They emulate the observed mean Hubble residuals measured as functions of supernova colour, in two separate bins of the host galaxy 
stellar mass. The mean difference between $\delta$ from the two branches corresponds to the classic (achromatic) mass-step correction, 
which turns out to provide only a partial description (no colour dependence) of the observations. The relatively small scatter in the $\delta$ values estimated for 
individual supernovae in the Hubble flow demonstrates that survey selection effects contribute to $\delta$ as second-order corrections.

\subsection{Dust model}\label{dust_model}

\begin{table*}
\begin{center}
\begin{tabular}{lcccccc}
\hline
dust model & \multicolumn{2}{c}{Pantheon+ \citep[][P23]{Popovic2021}} & \multicolumn{4}{c}{this work}  \\
 & & & & \\
applies to & \multicolumn{2}{c}{Calibration and Hubble Flow} & \multicolumn{2}{c}{Calibration} & \multicolumn{2}{c}{Hubble flow} \\
\hline
stellar mass bin  & $M_{\star}>10^{10}M_{\odot}$ & $M_{\star}<10^{10}M_{\odot}$ &  $M_{\star}>10^{10}M_{\odot}$ & $M_{\star}<10^{10}M_{\odot}$ &  $M_{\star}>10^{10}M_{\odot}$ & $M_{\star}<10^{10}M_{\odot}$\\
\hline
$\langle c_{\rm int}\rangle$ & $-0.077$ & $-0.077$ &  \multicolumn{2}{c}{the same as P23} &  \multicolumn{2}{c}{the same as P23} \\
$\sigma_{c_{\rm int}}$ & 0.058 & 0.058 & \multicolumn{2}{c}{the same as P23} &  \multicolumn{2}{c}{the same as P23}  \\
$\langle \beta_{\rm SN}\rangle$ & 2.064 & 2.064 & \multicolumn{2}{c}{the same as P23} &  \multicolumn{2}{c}{the same as P23}  \\
$\sigma_{\beta_{\rm SN}}$ & 0.308 & 0.308 & \multicolumn{2}{c}{the same as P23} &  \multicolumn{2}{c}{the same as P23}  \\
$\langle R_{\rm B} \rangle$ & 3.138 & 4.026 & $\mathbf{4.3}$ & $\mathbf{4.3}$ &  \multicolumn{2}{c}{the same as P23}  \\
$\sigma_{R_{\rm B}}$ & 1.061 & 1.481 & $\mathbf{0.4}$ & $\mathbf{0.4}$ &  \multicolumn{2}{c}{the same as P23}  \\
$\langle E(B-V)\rangle$ & 0.11 & 0.087 & \multicolumn{2}{c}{the same as P23}  &  \multicolumn{2}{c}{the same as P23}  \\
$\gamma$ & 1 & 1 & $\mathbf{3.44}$ & $\mathbf{3.44}$ &  \multicolumn{2}{c}{the same as P23}  \\
\hline
\end{tabular}
\caption{Hyperparameters of the prior probability distributions used to simulate observed light curve parameters 
of type Ia supernovae (peak magnitude $m_{\rm B}$ and colour parameter $c$) through physically 
motivated latent variables, i.e. intrinsic colour $c_{\rm int}$, dust reddening $E(B-V)$, 
the linear coefficient $\beta_{\rm SN}$ of the $m_{\rm B}-c_{\rm int}$ relation, and the total-to-selective extinction 
coefficient $R_{\rm B}$ in the $B$-band. The simulated light curve parameters are used to estimate the bias correction 
of supernova peak magnitudes and the additive model scatter, with the primary purpose of probabilistic modelling of 
extinction in supernova host galaxies \citep[for details see 
section~\ref{dust_model} or][]{Popovic2021}. 
The left columns show the parameters of the P23 model which were used to determine biases and scatter in the Pantheon+ supernovae, 
both in the calibration galaxies and the Hubble flow. The right columns show the new model proposed in this work 
(see section~\ref{new_dust} for explanation). The proposed modifications (boldfaced values) apply solely to supernovae in the calibration galaxies. 
Except for dust reddening, all latent variables are modelled using prior Gaussian 
distributions. The table lists their mean values ($\mu_{\rm X}$) and standard deviations ($\sigma_{\rm X}$). The 
distribution of dust reddening follows a $\gamma$ distribution model (equation~\ref{pdf_gamma}). The table shows the shape parameter $\gamma$ (with an exponential 
distribution from the P23 model recovered for $\gamma=1$) and the mean dust reddening $\langle E(B-V)\rangle$. 
}\label{dust-models}
\end{center}
\end{table*}

The dust model adopted for analysis of type Ia supernovae from the Pantheon+ catalogue was introduced by 
\citet{Brout2021} and the final constraints on the model parameters were presented in \citet{Popovic2021}. The model 
assumes that the colour parameter $c$ (in SALT2, chosen in a way that it resembles closely the time-independent component of the apparent rest-frame $B-V$ colour) is a sum of two independent latent variables: the supernova intrinsic colour 
$c_{\rm int}$ and the dust reddening $E(B-V)$, i.e.
\begin{equation}
c=c_{\rm int}+E(B-V).
\label{c}
\end{equation}
The resulting supernova brightness depends both on $c_{\rm int}$ and $E(B-V)$. The model employs linear relations 
for both variables such that the change in $m_{\rm B}$ is given by
\begin{equation}
\Delta m_{\rm B}=\beta_{\rm SN}c_{\rm int}+R_{\rm B}E(B-V),
\label{Deltamb}
\end{equation}
where $\beta_{\rm SN}$ is the slope of the relation between the unextinguished peak magnitude and the intrinsic supernova colour $c_{\rm int}$, and $R_{\rm B}$ is the total-to-selective extinction coefficient in the $B$-band. Equations~(\ref{c}--\ref{Deltamb}) contain four latent 
variables $\{c_{\rm int},E(B-V),\beta_{\rm SN},R_{\rm B}\}$ whose values vary between individual supernovae. The statistical properties of these variations can be observationally constrained by 
means of modelling distributions of supernova colour parameters $c$ and Hubble residuals $\Delta m_{\rm B}$. The results of this modelling 
depend on the choice of the assumed probability distribution for each latent variable. The P23 dust model assumed that the dust reddening follows an exponential distribution, i.e.
\begin{equation}
p(E(B-V))=\exp\Big(-\frac{E(B-V)}{\tau}\Big)\frac{1}{\tau},
\label{ebv_exp}
\end{equation}
where $\tau$ is a free parameter (hyperparameter) and $E(B-V)>0$. The three remaining latent variables were modelled using Gaussian probability distributions whose means and variances were free hyperparameters of the Bayesian hierarchical model. The exponential model adopted for $E(B-V)$ is the simplest proposal distribution of a positively defined variable. It is the most likely functional form 
when the only constraint derivable from data is the mean value. Unphysical values of $R_{\rm B}$ were excluded by imposing a lower limit equal to $1.5$ \citep{Brout2021}. In addition, the distributions of $E(B-V)$ and $R_{\rm B}$ were fitted independently 
in the two stellar mass bins of supernova host galaxies.

The statistical properties of the observed supernova colour parameters $c$ place constraints on population properties of $c_{\rm int}$ and $E(B-V)$, 
while supernova Hubble residuals as a function of $c$ constrain the coefficients $\beta_{\rm SN}$ and $R_{\rm B}$. The apparent difference between 
colour-dependent Hubble residuals in the high and low stellar-mass host galaxies are ascribed to a difference between the mean values of the extinction 
coefficient $R_{\rm B}$, while the colour-dependent scatter is modelled as a scatter in individual $R_{\rm B}$ values. Table~\ref{dust-models} 
provides a summary of the model and best fit parameters (hyperparameters) obtained by \citet{Popovic2021}.

\begin{figure*}

    \begin{subfigure}[t]{0.45\textwidth}
        \centering
        \includegraphics[width=\linewidth]{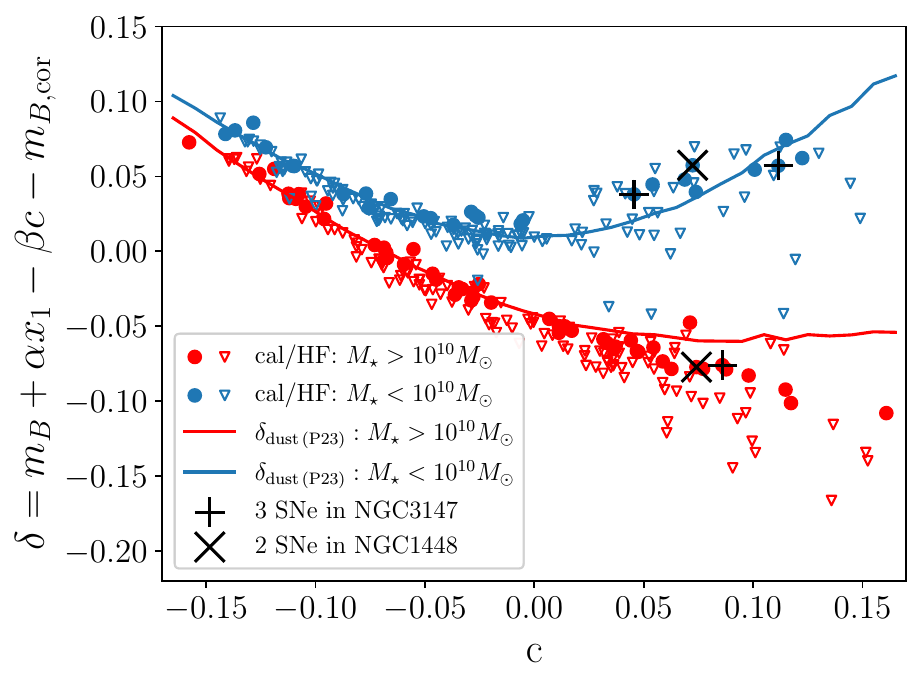} 
        \caption{} \label{HD_ori_b}
    \end{subfigure}
    \begin{subfigure}[t]{0.45\textwidth}
    \centering
        \includegraphics[width=0.97\linewidth]{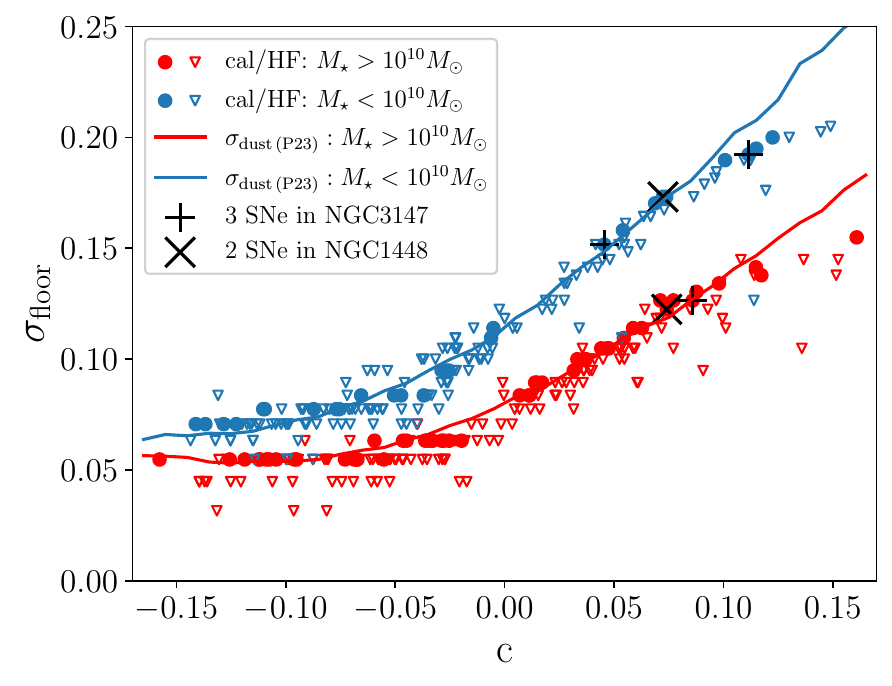} 
        \caption{} \label{HD_ori_c}
            \end{subfigure}

    \centering
    \begin{subfigure}[l]{0.85\textwidth}
        \centering
        \includegraphics[width=\linewidth]{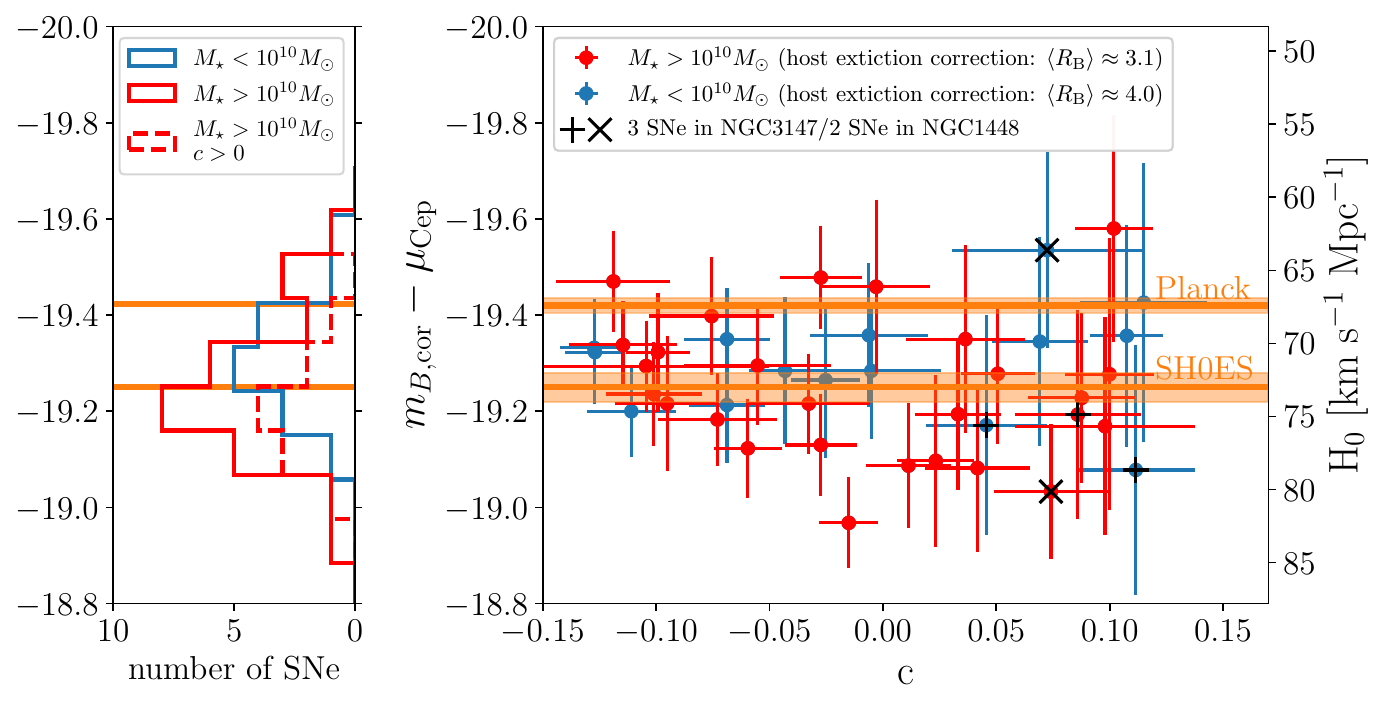} 
        \caption{} \label{HD_ori_a}
    \end{subfigure}

    \caption{\textit{Panels (a) and (b):} The bias correction of supernova peak magnitudes 
    (\textit{left}) and the floor uncertainty $\sigma_{\rm floor}$ (the intrinsic scatter model) (\textit{right}) estimated for each individual supernova (including duplicates) from the 
    Pantheon+ compilation, in the calibration sample (cal) and the Hubble flow (HF). Both the bias and the scatter are primarily driven by a model of supernova intrinsic colours, dust 
    reddening and extinction, developed by \citet{Popovic2021} and employed in the Pantheon+ compilation. The solid curves show the P23 model 
    predictions based on Monte Carlo simulations of supernova properties with the best fit parameters from \citet{Popovic2021}. 
    The apparent differences between the high and low stellar-mass host galaxies are 
    primarily driven by: different mean values of the total-to-selective extinction coefficient $\langle R_{\rm B}\rangle$ 
    (resulting in different bias profiles; see Table~\ref{dust-models}) and different degrees of scatter in individual $R_{\rm B}$ values per sight line 
    (resulting in different $\sigma_{\rm floor}(c)$ profiles; see Table~\ref{dust-models}). The P23 model is trained on $z>0.03$ supernovae and extrapolated to the calibration sample. 
    \textit{Panel (c):} \textit{Right}: Supernova corrected magnitudes (Pantheon+) in the calibration galaxies (measurements from duplicates combined into one), compared to distance moduli derived from 
    the Cepheid observations (SH0ES), as a function of supernova colour parameter $c$. The left y-axis shows the absolute magnitude $M_{\rm B}$ (since there are no K corrections for the nearby supernovae in the calibration sample), while 
    the right axis shows the corresponding $H_{0}$ (equation~\ref{M_B_H_0}). 
    Red supernovae ($c>0$) in the high stellar-mass hosts ($M_{\star}>10^{10}M_{\odot}$) tend to be intrinsically fainter implying higher values 
    of the Hubble constant. The horizontal lines (shaded bands) show the Hubble constant values (uncertainties) from SH0ES 
    \citep[][matching the best fit $M_{\rm B}$ derived from the calibration data]{Riess2022} and Planck \citep{Planck2020_cosmo}. The black symbols on all panels show two sets of supernova siblings for which the bias and scatter estimations follow both the high and low stellar-mass solutions 
    due to discrepant stellar mass entries assigned to each sibling originating from the same host. \textit{Left}: The adjacent panel shows the histograms of supernova absolute magnitudes derived from Cepheid distance moduli ($m_{\rm B,corr}-\mu_{\rm Cep}$) in the high and low 
    stellar-mass host galaxies. Type Ia supernovae in the high stellar-mass host galaxies are systematically fainter than those in the low stellar-mass hosts.}
    \label{HD_ori}
\end{figure*}

Using the P23 model \citep[][see also Table~\ref{dust-models}]{Popovic2021} we compute the bias $\delta_{\rm dust}(c)$ as the difference between 
$m_{\rm B}(c)$ predicted by the model and the linear approximation given by the Tripp correction, as defined by equation~(\ref{mb_main}). 
We do not include any selection effects and assume negligible errors in supernova light curve parameters, hence the bias solely quantifies the non-linearity 
of the $m_{\rm B}(c)$ relation due to dust. The computation is based on generating a Monte Carlo sample of supernovae with latent variables drawn from the 
dust model distributions and obtaining observed colours $c$ given by equation~(\ref{c}) and supernova magnitudes 
(at an arbitrary fixed redshift and an arbitrary absolute magnitude) given by equation~(\ref{Deltamb}). We assume equal fractions of high and low stellar-mass host galaxies, consistently with what we find in the Hubble flow sample 
used by \citet{Riess2022}. The bias is estimated in bins of colour parameter $c$ and host galaxy stellar mass in the same way as in \citet{Brout2021} and 
\citet{Popovic2021}. It is the mean residual magnitude $m_{\rm B}$ with respect to the best fit Tripp model obtained from fitting the model to all simulated supernovae 
in both bins of host galaxy stellar mass, i.e.
\begin{equation}
\delta_{\rm dust\,(P23)}(c,M_{\star})=\langle \Delta m_{\rm B,dust\,(P23)}(c,M_{\star})-\beta c-\Delta_{0}\rangle,
\label{bias_gen}
\end{equation}
where the subscript 'P23' indicates the dust model from \citet{Popovic2021} employed in the simulation. The dependence on stellar mass is realized via the step function defining the stellar mass bins 
and $\{\beta,\Delta_{0}\}$ are free parameters of the 
Tripp model: the slope and the offset at $c=0$ of a linear relation 
between $\Delta m_{\rm B,dust\,(P23)}$ and $c$. The best-fit slope of the Tripp model found within the range of observed colour parameters is $\beta=3.0$ which is consistent with 
the estimate from the full forward modelling of type Ia supernovae \citep{Brout2022}. The parameter $\beta$ quantifies an average slope of the $\Delta m_{\rm B,dust\,(P23)}-c$ relation which is driven by $\beta_{\rm SN}$ at the blue-colour end and by $R_{\rm B}$ at the red-colour end of the $c$ distribution. The offset $\Delta_{0}$ includes average corrections due to intrinsic colours and dust reddening at $c=0$.

Figure~\ref{HD_ori_b} shows that the computed 
bias $\delta_{\rm dust\,(P23)}$ closely reproduces the actual bias $\delta$ from the Pantheon+ catalogue. This demonstrates that the dust model is 
the leading component of all corrections encapsulated in $\delta$, in both the calibration and Hubble flow samples. The colour-dependent slope of $\delta_{\rm dust\,(P23)}(c)$ 
changes from $\langle \beta_{\rm SN}\rangle-\beta\approx -1$ for the bluest colours ($c\approx -0.15$), where the change of the apparent magnitude is dominated by 
the intrinsic colour correction, to $\langle R_{\rm B}\rangle -\beta\approx 1$ for low stellar-mass host galaxies (or $\langle R_{\rm B}\rangle -\beta\approx 0$ for 
high stellar-mass host galaxies) for the reddest supernovae ($c\approx 0.15$) for which the colour dependence is driven by extinction. The apparent difference 
between $\delta_{\rm dust\,(P23)}(c)$ in the two stellar mass bins of supernova host galaxies is primarily driven by the different mean $R_{\rm B}$ coefficients assigned to these bins. 
This colour-dependent difference was shown to improve a simple colour-independent mass-step correction in supernova absolute magnitudes \citep{Popovic2021,Brout2021}.

\subsection{Covariance and intrinsic scatter model}

The covariance matrix in the calibration sample is the sum of statistical and systematic covariances 
of corrected supernova magnitudes and distance moduli from Cepheid observations 
\citep[$\mathsf{C}_{\rm stat+syst}^{\rm SN}$ and $\mathsf{C}_{\rm stat+syst}^{\rm Cepheid}$, respectively, in][]{Brout2022}. 
The statistical covariance matrix of supernova corrected magnitudes contains diagonal elements 
given by
\begin{equation}
\sigma_{\rm tot}^{2}=\sigma_{\rm meas}^{2}+\sigma_{\rm floor}^{2}(c,M_{\star}),
\end{equation}
where $\sigma_{\rm meas}$ is the measurement uncertainty of the peak magnitude corrected with the Tripp 
model and $\sigma_{\rm floor}$ is a colour- and stellar mass-dependent floor uncertainty describing intrinsic scatter in supernova Hubble residuals. The floor uncertainty is computed for the best fit model of dust and supernova intrinsic colours adopted in the Pantheon+ supernova compilation, where the dependence on stellar mass is realized via the step function defining the stellar mass bins \citep{Brout2022}.\footnote{We note that the factor $f$ used to scale the contribution from $\sigma_{\rm meas}^{2}$ in equation (3) of \citet[][]{Brout2022} 
is equal to 1 both in the calibration sample and the Hubble flow, while the colour-independent intrinsic scatter 
\citep[$\sigma_{\rm grey}^{2}$ in equation (4) of][]{Brout2022} vanishes for the dust model adopted in the Pantheon+ supernova compilation.}
Projections of the covariance matrix elements 
onto the corrected peak magnitude uncertainties are computed assuming standard best-fit coefficients of the Tripp model 
\citep[see e.g.][]{Kessler2017,Wojtak2022} with $\beta\simeq 3.0$ \citep{Brout2022}. The only non-vanishing 
non-diagonal elements of the statistical covariance matrix are those associated with supernova duplicates 
which represent independent measurements of light curve parameters of the same supernova. These elements are equal 
to $\sigma_{\rm floor}^{2}(c,M_{\star})$.

Figure~\ref{HD_ori_c} shows the floor uncertainty $\sigma_{\rm floor}(c)$ (square root of $biasCor\_m\_b\_COVADD$
in the Pantheon+ catalogue) for supernovae in the calibration sample and the Hubble flow. The uncertainty as a function 
of colour is well reproduced by scatter arising solely from the P23 model \citep{Popovic2021} employed in the Pantheon+ data 
analysis. We show the model's prediction with solid curves. They are computed 
using Monte Carlo simulations of supernova apparent magnitudes $\Delta m_{\rm B,dust(Pop23)}$ and colour parameters 
$c_{\rm dust\,(Pop23)}$, as outlined in section~\ref{dust_model}. 
Following \citet{Brout2021} and \citet{Popovic2021}, the floor uncertainty is given by the standard deviation of the 
simulated apparent magnitudes, as a function of colour $c$, in the two separate stellar mass bins, i.e.
\begin{eqnarray}
    \sigma_{\rm dust\,(Pop23)}(c,M_{\star})&=&\langle [\Delta m_{\rm B,dust\,(Pop23)}(c,M_{\star})\nonumber\\ 
    &-& \langle \Delta m_{\rm B,dust\,(Pop23)}(c,M_{\star})\rangle]^{2}\rangle^{1/2}.
\end{eqnarray}
The computed values of $\sigma_{\rm dust\,(Pop23)}$, given the available information on the P23 model \citep{Popovic2021}, are rescaled by 0.83 in order to match the corresponding floor uncertainties provided in the catalogue.

Figure~\ref{HD_ori_c} demonstrates a substantial difference between the high and low stellar-mass host galaxies for red 
supernova colours ($c\gtrsim 0$). This effect is primarily driven by different values of scatter in $R_{\rm B}$ assigned to the stellar mass bins 
in the P23 model ($\sim 50$ per cent larger scatter in the low stellar-mass host galaxies; see Table~\ref{dust-models}). It reflects the actual 
dependence of scatter in supernova Hubble residuals on colour and host galaxy stellar mass. It is also worth noting that the floor uncertainty provides a 
substantial contribution to the diagonal elements of the joint covariance matrix $\mathsf{C}_{\rm stat+syst}^{\rm SN+Cepheid}=\mathsf{C}_{\rm stat+syst}^{\rm SN}+\mathsf{C}_{\rm stat+syst}^{\rm Cepheid}$: on average $25$ per cent at $c<0$ and $40$ per cent at $c>0$.

\subsection{Assumptions and caveats}\label{assumptions}

The model of dust and supernova intrinsic colours developed by \citet{Popovic2021} and applied to the Pantheon+ supernova 
data to estimate the bias $\delta(c,M_{\rm star})$ and the floor uncertainty $\sigma_{\rm floor}(c,M_{\rm star})$ was fitted to 
all supernovae at redshifts $z>0.03$. Strictly speaking, the model describes population properties of dust reddening and 
extinction solely in supernova host galaxies of the training data set. Applying it to the calibration galaxies, which are different 
from those in the training data due to selection bias, is an extrapolation which should be justified. Below we discuss some 
concerns regarding this extrapolation.

\begin{figure}
    \centering
        \includegraphics[width=\linewidth]{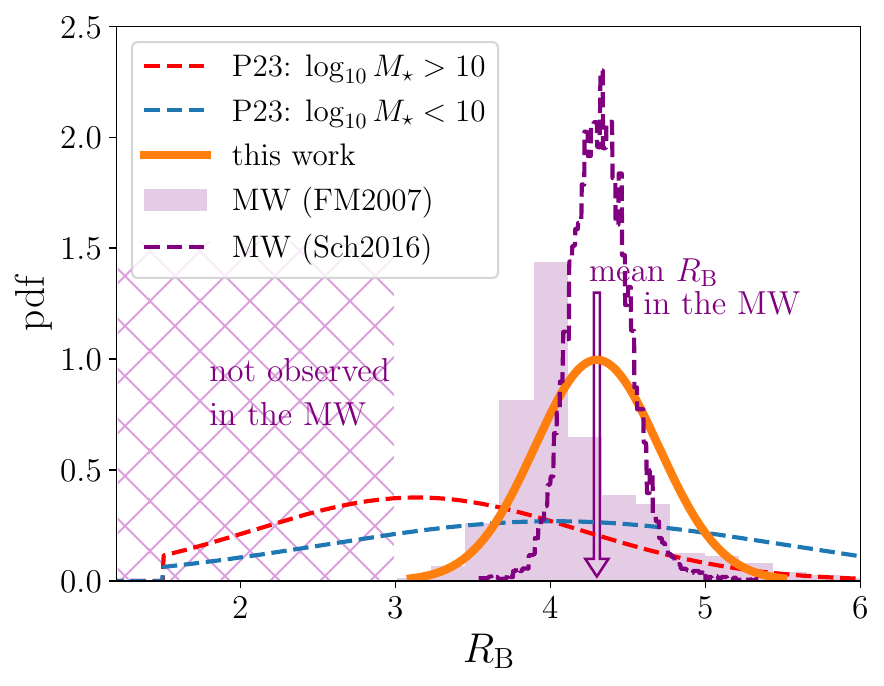} 
    \caption{Model distributions of the total-to-selective extinction coefficient $R_{\rm B}$ used to correct supernova peak magnitudes 
    for extinction in the calibration galaxies. The dashed (red and blue) curves show the P23 \citep{Popovic2021} model adopted in 
    the SH0ES measurement of the Hubble constant \citep{Riess2022}, while the solid orange curve shows the model used in this work. 
    The models are compared to observational constraints on $R_{\rm B}\approx R_{\rm V}+1$ in the Milky Way (MW) from 
    \citet[][FM2007]{Fitzpatrick2007} and \citet[][Sch2016]{Schlafly2016}. The mean MW value indicated by the arrow is also adopted in the extinction model used for colour corrections of Cepheids \citep[baseline model of][]{Riess2022}.
    }\label{RB_models}
\end{figure}

\begin{itemize}
\item \textit{Does the mass-step correction ascribed to different mean values of $R_{\rm B}$ in the corresponding stellar mass bins 
of supernova host galaxies apply to late-type galaxies in the calibration sample?} Historically, the mass-step correction was 
found in large samples of supernovae with host galaxies encompassing all morphological types \citep[see e.g.][]{Kelly2010}. 
In this respect, the analysis of \citet{Popovic2021} is not different in a sense that the resulting difference between $\langle R_{\rm B}\rangle$ 
in low and high stellar-mass host galaxies is implicitly averaged over all observed types of supernova host galaxies. Whether the same mass-step correction holds 
across galaxy morphological types, in particular those represented by the calibration galaxies, is an empirical question which has not yet been addressed. 
Moreover,
there is no plausible explanation of a physical mechanism that can account for an abrupt change of the total-to-selective 
mean extinction coefficient at $M_{\star}=10^{10}M_{\odot}$. 
Although some non-standard models of scattering predicting low values of 
$R_{\rm B}$ have been hypothesised \citep{Goobar2008}, it is not clear why they would manifest themselves solely in the high stellar-mass hosts.
In fact, the mass-step in $\langle R_{\rm B}\rangle$ may be an emerging property 
rather than a genuine signature of different extinction laws in galaxies with stellar masses above and below $10^{10}M_{\odot}$ 
(see our comments in the Introduction). 

\item \textit{How can one reconcile extinction corrections of supernovae (with $\langle R_{\rm B}\rangle=3.1$) and Cepheids 
(with $R_{\rm B}=4.3$) in the same high stellar-mass calibration galaxies ($M_{\star}>10^{10}M_{\odot}$, about 67 per cent of all 
calibration galaxies)?} 
The colour correction of Cepheids in the baseline model of the SH0ES measurement was computed assuming Milky Way-like extinction curve 
from \citet{Fitzpatrick1999} with $R_{\rm B}\approx R_{\rm V}+1=4.3$ \citep{Riess2022}. It was demonstrated that plausible variations 
in these assumptions have negligible impact on the effective total-to-selective extinction coefficient in the infrared (F160W) and thus the colour corrections 
of Cepheids. While the baseline extinction model for Cepheids seems to be the most natural choice for estimating the extinction correction for Cepheids, 
the P23 model assumes a different extinction model with $\langle R_{\rm B}\rangle=3.1$ towards high stellar-mass host galaxies. This would require a complete spatial separation between Cepheids and the supernovae, and the associated dust properties along their 
sight lines. However, both supernova and host galaxy properties strongly suggest that the calibration supernovae should coincide closely with the disk component 
containing Cepheids. Firstly, the calibration supernova sample consists mainly of high-stretch (slowly declining light curves) supernovae which are known to originate 
predominantly from young and star forming environments, common to stellar disk components \citep{Sullivan2006,Rigault2013}. Low-stretch supernovae with $x_{1}\lesssim -0.8$, which 
make up about 30 per cent of all type Ia supernovae in the Hubble flow and are found in the lowest local specific star-formation regions \citep[with $\log_{10}({\rm sSFR})\lesssim -11$, see][]{Rigault2020}, are 3 times less numerous in the calibration galaxies than in the Hubble flow. Secondly, the calibration galaxies are disk-dominated systems 
\citep[see Figure~4 of ][]{Riess2022}. For this class of host galaxies, type Ia supernovae trace closely the light distribution of stellar disks \citep{Prichet2024}.

\item \textit{How can one reconcile $\langle R_{\rm B}\rangle=3.1$ assumed by the P23 model for the high stellar-mass host galaxies 
with independent estimates of extinction coefficients in similar late-type galaxies?}
The adopted mean extinction coefficient is not only substantially lower than the average $R_{\rm B}$ measured in the Milky Way, but furthermore it can barely match the lowest 
values found in individual sight lines  \citep[see e.g.][see also Figure~\ref{RB_models}]{Fitzpatrick2007,Maiz2024}. It is also significantly lower than typical $R_{\rm B}$ measured in 
similar star-forming galaxies \citep{Salim2018}. $50$ per cent of $R_{\rm B}\lesssim 3$ values predicted by the model distribution 
are not observed in the Milky Way \citep{Draine2003}. Constraints on $R_{\rm B}$ from multi-band observations of type Ia supernovae depend heavily on models employed to separate dust reddening from 
supernova intrinsic colours. Based on different methodologies and supernova samples, the mean $R_{\rm B}$ is typically found within a range between 3 \citep[see e.g.][]{Nobili2008,Burns2014}
and 4 \citep[see e.g.][]{Thorp2021,Ward2023,Wojtak2023} \citep[for a comprehensive compilation of recent results, see Table~A1 of][]{Thorp2024}. Although these estimates are typically below the mean $R_{\rm B}$ measured in the Milky Way, 
a substantial scatter in $R_{\rm B}$ (with $\sigma_{\rm R_{\rm B}}\gtrsim 0.6$ found consistently in all studies) implies that Milky Way-like extinction coefficients lie well within 
the most probable values of the $R_{\rm B}$ distribution inferred from supernova data. This fact is well illustrated by Figure~\ref{RB_models} for the $R_{\rm B}$ distributions 
from the P23 model, where the Milky Way-like reference $R_{\rm B}$ is only $1.3\sigma_{R_{\rm B}}$ higher than the mean value found in the high-stellar mass host galaxies. 
Based on these statistical properties, we can expect that any random supernova sample contains a subpopulation of host galaxies or supernova sight lines 
for which Milky Way-like extinction is observed. In our work, we assume that the calibration galaxies, which may be seen as analogues of the Milky Way, give rise to Milky Way-like extinction along the supernova sight lines.

\end{itemize}

In an attempt to minimise selection bias related to apparent differences between host galaxies in the calibration sample and the Hubble flow, \citet{Riess2022} narrowed down 
the Hubble flow data to late-type host galaxies, based on morphological type assessed from the best available optical images. They showed that the mean 
specific star formation rate of supernova host galaxies in the Hubble flow matches that of the calibration galaxies. Furthermore, supernovae in the Hubble flow 
are selected in the same range of light curve parameters as those in the calibration sample. \citet{Riess2022} argued that these selections rule out any significant unaccounted for differences in extinction and showed that they have negligible impact on the Hubble constant estimation. The important question 
here is if the expected match is sufficiently close. The main concern lies in the fact that the matching conditions involve global and rather simplified galaxy properties, while an exact match of the population properties of the extinction may require probing local supernova environments. Secondly, employing identical cuts in supernova colour parameters does not guarantee that the relative effects of extinction and intrinsic colours are the same in both supernova samples.  Assuming conservatively that a close match of global specific average star formation rates in the calibration sample and the Hubble flow, as obtained by \citet{Riess2022}, implies equal average dust reddening (average dust column density in each sample), one cannot rule out possible differences between higher moments (higher order corrections) of the reddening distributions. We show in section~\ref{new_dust} that this effect can have a strong impact on the accuracy of propagating distance measurement from the calibration sample to the Hubble flow.

\section{Consistency tests}\label{consistency}

We begin by showing the most essential part of the calibration data. Figure~\ref{HD_ori_a} compares corrected supernova peak magnitudes 
to distance moduli from Cepheids, as a function of supernova colour parameter $c$. For the sake of readability, we show the data 
for 42 distinct supernovae rather than 77 independent measurements of light curve parameters (including duplicates). 
The best fit peak magnitude and its total uncertainty, as well as the measurements of Cepheid distance moduli, are taken 
from Table~2 of \citet{Riess2022}. The plotted error bars in the magnitudes include uncertainties both in $m_{\rm B,corr}$ and $\mu_{\rm Cep}$, 
i.e.\ $\sigma=(\sigma_{m_{\rm B,corr}}^{2}+\sigma_{\mu_{\rm Cep}}^{2})^{1/2}$. Supernova colour parameters and their uncertainties 
are computed by combining measurements from duplicates associated with each distinct supernova. We assume that the combined measurement 
is given by a product of Gaussian distributions of light curve parameters with the mean values and covariance matrix provided in the Pantheon+ catalogue.

Our analysis is focused on the calibration sample which places constraints on the absolute magnitude $M_{\rm B}$ 
of type Ia supernovae. This is what is shown on the left axis of Figure~\ref{HD_ori_a}. In order to see a direct effect of any 
change in $M_{\rm B}$ on the Hubble constant, we will use the following mapping between $M_{\rm B}$ and $H_{0}$:
\begin{equation}
M_{\rm B}+19.25=5\log_{10}(h/0.73),
\label{M_B_H_0}
\end{equation}
consistent with the best fit baseline model of the SH0ES measurement \citep{Riess2022}. The mapping is entirely and independently 
constrained by supernovae in the Hubble flow with a precision of about $0.009$~mag (compared to the $\sim0.03$~mag precision in measuring 
$M_{\rm B}$ in the calibration sample). We use it to show the expected Hubble constant corresponding to different possible 
values of $M_{\rm B}$ found in the calibration sample. This is what is shown on the right axis of Figure~\ref{HD_ori_a}. We 
note, however, that our final Hubble constant measurement presented in section~\ref{final_H0} is based on a complete analysis 
of data in the calibration sample and the Hubble flow.

\subsection{Supernovae in high stellar-mass host galaxies}\label{tests_ori}

\begin{figure*}
    \centering
    \begin{subfigure}[l]{0.45\textwidth}
        \centering
        \includegraphics[width=\linewidth]{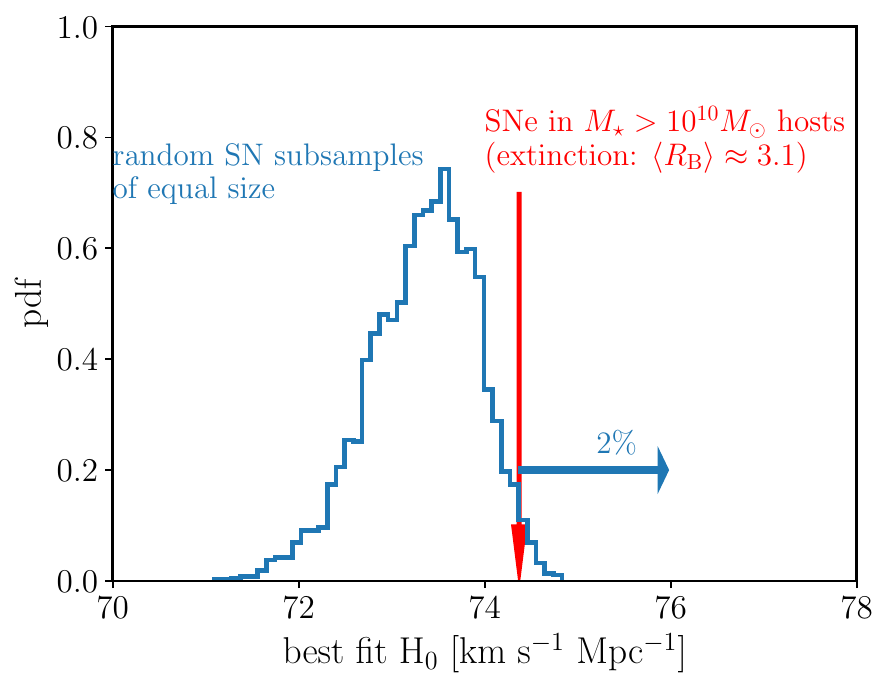} 
        \caption{} \label{anoma_ori_a}
    \end{subfigure}
    \begin{subfigure}[r]{0.45\textwidth}
        \centering
        \includegraphics[width=\linewidth]{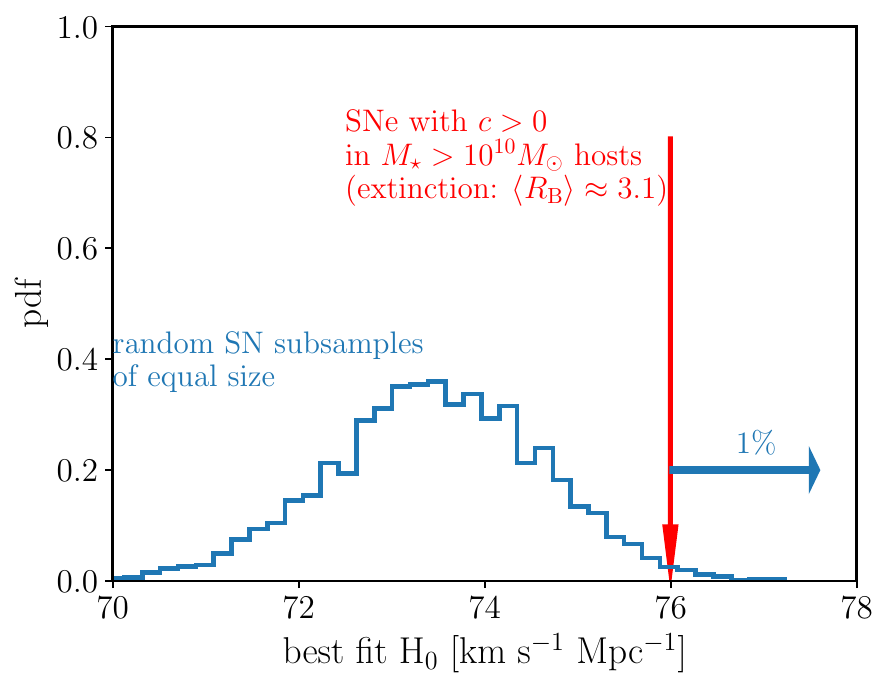} 
        \caption{} \label{anoma_ori_b}
    \end{subfigure}
    \caption{Tests of intrinsic consistency in the calibration data. The arrows indicate the best fit $H_{0}$ 
    (derived from the best fit $M_{\rm B}$ and using equation~(\ref{M_B_H_0})) 
    obtained for supernovae in the high stellar-mass ($M_{\star}>10^{10}M_{\odot}$, with extinction correction assuming 
    $\langle R_{\rm B}\rangle\approx 3.1$) host galaxies (\textit{left}) and its subsample with red supernovae ($c>0$) 
    (\textit{right}). The blue distributions show the 
    corresponding best fit values obtained from random control samples with the same number of distinct supernovae. 
    Supernovae in the high stellar-mass category give rise to systematically higher $H_{0}$ (lower intrinsic luminosity) than 
    the corresponding control samples. The best fit values are $2.0\sigma$ (\textit{left}) and $2.3\sigma$ (\textit{right}) 
    outliers.
    }
\label{anoma_ori}
\end{figure*}

If the dust model employed in the Pantheon+ supernova compilation provides an equally accurate description of 
dust reddening and extinction in the calibration sample as in the Hubble flow, one should not expect 
any significant differences between $M_{\rm B}$ measured independently from high and low stellar-mass host 
galaxies, or across supernova colours. However, supernova absolute magnitudes derived from Cepheid 
distance moduli appear to be fainter in the high stellar-mass host galaxies, as is demonstrated by the histograms 
of $M_{\rm B}$ in Figure~\ref{HD_ori_a}. We test the signifiance of this apparent offset by comparing best fit $M_{\rm B}$ inferred from a subsample 
of supernovae in $M_{\star}>10^{10}M_{\odot}$ host galaxies (extinction correction with $\langle R_{\rm B}\rangle=3.1$) 
to fits using random control subsamples of equal size.

We use the standard likelihood $L$ of the Pantheon+ data set \citep{Brout2022} restricted to the calibration 
data and given by
\begin{equation}
\ln L \propto -\frac{1}{2} \pmb{\Delta\mu^{T}}(\mathsf{C}_{\rm stat+syst}^{\rm SN+Cepheid_{cal}})^{-1}\pmb{\Delta\mu}-\frac{1}{2}\ln\det(\mathsf{C}_{\rm stat+syst}^{\rm SN+Cepheid_{cal}})
\end{equation}
where $\mathsf{C}_{\rm stat+syst}^{\rm SN+Cepheid_{cal}}$ is a 77-dimensional covariance matrix of all supernova duplicates 
in the calibration sample (extracted from the complete matrix $\mathsf{C}_{\rm stat+syst}^{\rm SN+Cepheid}$ provided in 
the Pantheon+ data release), $\pmb{\Delta\mu}$ is a vector with the $i$-th component given by 
$\Delta\mu_{i}=m_{\rm B,corr,i}-M_{\rm B}-\mu_{\rm Cep,i}$, and $M_{\rm B}$ (or equivalently $H_{0}$ from equation~(\ref{M_B_H_0})) is the only free parameter in the fit. We note that the normalisation factor given by $\det(\mathsf{C}_{\rm stat+syst})$ depends on the extinction model. Including it is relevant for model selection based on comparing maximum likelihood values (see section~\ref{observational_tests}). We compute the best fit model parameter using a Monte Carlo Markov Chain technique implemented in the \textit{emcee} code \citep{emcee}. All results are provided 
in the form of posterior means for a flat prior distribution in $H_{0}$. As expected based on the baseline results from \citet{Riess2022}, fitting the full calibration data set yields $M_{\rm B}=-19.25\pm0.03$ or equivalently $H_{0}=73.03\pm1.00$~km~s$^{-1}$~Mpc$^{-1}$.

Figure~\ref{anoma_ori} shows the best fit $H_{0}$ obtained for two subsamples of the calibration supernovae: supernovae 
in high stellar-mass host galaxies (Figure~\ref{anoma_ori_a}) and a subsample further restricted to include only red supernovae with $c>0$ for 
each duplicate (Figure~\ref{anoma_ori_b}). They are compared to analogous best fit parameters 
obtained for random control subsamples containing the same number of distinct supernovae. The results demonstrate 
that supernovae in the high stellar-mass host galaxies are systematically fainter (with systematically higher derived values of the Hubble constant). 
The effect is particularly strong for red supernovae, as is evident from Figure~\ref{HD_ori_a}. 
The best fit $H_{0}$ values lie in the 1--2 per cent range (equivalent to $2.3\sigma$ and $2.0\sigma$ outliers, where the confidence levels are given by two-sided intervals of the Gaussian distribution) of the highest estimates derived from random control subsamples. The analogous best-fit value obtained for supernovae in the low stellar-mass 
host galaxies is consistent with the distribution of random control subsamples. However, it is shifted towards brighter 
luminosities ($-0.76\sigma$ shift from the mean) implying a $2.7\sigma$ intrinsic tension between supernovae in the two stellar mass bins. 

Systematically lower luminosities of type Ia supernovae in the high stellar-mass host galaxies were also found by \citet{Riess2022}. Since their analysis is based on a joint fit to the Cepheid and supernova data, involving distance moduli of the calibration galaxies as latent variables, it is expected that a part of the signal is absorbed by possible shifts in derived distance moduli. 
In contrast, our approach employs distance moduli fixed at the best fit values inferred from the Cepheid data alone and thus the effect of varying supernova samples can only impact the supernova absolute magnitude and $H_{0}$. We also note that the largest discrepancy found in our study occurs for red supernovae in the high stellar-mass hosts, which were not considered as a possible selection variant in \citet{Riess2022}.

Fits based solely on supernovae in high stellar-mass host galaxies favour lower intrinsic luminosities ($M_{\rm B}>-19.25$) 
and thus higher values of the Hubble constant than the reference fit based on the entire sample. This is visible both in 
Figure~\ref{HD_ori_a} and Figure~\ref{anoma_ori}. The change in the best fit $M_{\rm B}$ increases with supernova colour. 
For all supernovae in high stellar-mass galaxies we find $\Delta H_{0}=1.3$~km~s$^{-1}$~Mpc$^{-1}$ ($\Delta M_{\rm B}\approx 0.04$ mag), 
while for red supernovae ($c>0$) in the same stellar mass bin $\Delta H_{0}\approx 3$~km~s$^{-1}$~Mpc$^{-1}$ ($\Delta M_{\rm B}\approx 0.09$ mag). 
The colour dependence of $\Delta M_{\rm B}$ suggests that this could be an effect of underestimated extinction in the high stellar-mass host galaxies. 
This hypothesis is bolstered by realising that the apparent shift in the best fit $M_{\rm B}$ appears to coincide with the difference between the
bias correction of supernova peak magnitudes in the high and low stellar-mass hosts. Using
\begin{eqnarray}
\Delta_{\rm dust\,(P23)} & =\langle -\delta(c,M_{\star}>10^{10}M_{\odot})_{\rm dust\,(P23)}\nonumber\\
 & +\delta(c,M_{\star}<10^{10}M_{\odot})_{\rm dust\,(P23)}\rangle
\end{eqnarray}
with the model bias computed in section~\ref{dust_model} and shown in Figure~\ref{HD_ori_b}, we find $\Delta_{\rm dust(P23)}=0.065$ mag when averaging over 
all observed colours $|c|<0.15$ and $\Delta_{\rm dust(P23)}=0.099$ mag when averaging over red colours with $0<c<0.15$. A close match between the measured 
$\Delta M_{\rm B}$ and $\Delta_{\rm dust(P23)}$ signifies the apparent underestimation of extinction in the high stellar-mass hosts. Since the P23 model assumes nearly the same reddening distribution in all galaxies, this can be primarily ascribed to the difference between the mean extinction coefficients $\langle R_{\rm B}\rangle$ assigned to the stellar mass bins.

Repeating the test of consistency for 277 supernovae in the Hubble flow \citep[the baseline Hubble flow sample from][]{Riess2022} we do 
not find any noticeable changes in the best fit Hubble diagram normalisation. This is expected because the Hubble flow sample largely overlaps 
with the training data of the P23 model. Best fit normalisations obtained for supernovae in the high stellar-mass host galaxies agrees with those from random control 
subsamples within $0.5\sigma$ (and $1.4\sigma$ when restricting colours to $c>0$). The corresponding shift in $M_{\rm B}-5\log_{10}(h)$ is a mere $0.0037$~mag 
(and $0.02$~mag for $c>0$), or correspondingly $\Delta H_{0}=-0.13$~km~s$^{-1}$~Mpc$^{-1}$ ($\Delta H_{0}=-0.7$~km~s$^{-1}$~Mpc$^{-1}$ for $c>0$) 
for a fixed $M_{\rm B}$. On the one hand, small shifts are not surprising because the restricted Hubble flow from \citet{Riess2022} is 
not identical to the training data set used by \citet{Popovic2021}. On the other hand, it is perhaps surprising to see that late type host galaxies selected for 
the Hubble flow sample in \citet{Riess2022} resemble so closely the entire training sample \citet{Popovic2021}.

\subsection{Stellar mass estimates}

We find that two sets of supernova siblings are assigned discrepant stellar mass estimates in the Pantheon+ catalogue. 
For the NGC 3147 host galaxy (3 siblings), the stellar mass entries are $\log_{10}(M_{\star}/M_{\odot})=\{6.95,\,12.59\,,8.37\}$, and for 
NGC 1448 (2 siblings) $\log_{10}(M_{\star}/M_{\odot})=\{11.28,\,-9.0(\rm{N/A})\}$. As shown in Figure~\ref{HD_ori_b} and Figure~\ref{HD_ori_c}, 
discrepancies between the stellar mass entries propagate into the supernova peak magnitude corrections. The estimated biases 
and floor uncertainties of siblings are not estimated assuming consistently the same stellar mass bin of the dust model. Independent estimates 
of stellar masses from the literature place both host galaxies in the high stellar-mass bin \citep[see e.g.][]{Annuar2017,Biscardi2012,Thone2009}. 
This implies that it is three supernovae with low stellar-mass entries which were assigned with improper bias and scatter. We find that these 
supernovae have negligible impact on the consistency tests discussed in the above section. Redrawing random test samples omitting these three 
supernovae, the best fit $M_{\rm B}$ obtained from supernovae in the high stellar-mass host galaxies lies outside the $1.8\sigma$ range (and 
 $2.4\sigma$ range for red supernovae with $c>0$) of best fit values from the control samples.

\section{New dust model}\label{new_dust}

In this section, we describe a new dust model developed as a minimalistic modification of the original framework of \citet{Popovic2021}. 
The new model applies solely to the calibration galaxies. It is motivated by mitigating the intrinsic discrepancy between $M_{\rm B}$ derived 
from supernovae in high and low stellar-mass host galaxies and reconciling extinction laws used to correct Cepheid and supernovae magnitudes in $M_{\star}>10^{10}M_{\odot}$ hosts. 
The model entails modifications of the $R_{\rm B}$ distributions and the shape (2nd moment) of the $E(B-V)$ distributions adopted in the P23 model. 
It preserves all other properties of the P23 model, including: the distributions of intrinsic colours and $\beta_{\rm SN}$ coefficients, the slope of 
the colour correction in the Tripp formula and the mean reddening in each of the two stellar mass bins. The parameters of the new model are provided 
in Table~\ref{dust-models}.

Since our consistency tests do not reveal any unaccounted for residuals in the Hubble flow, we assume that the P23 model provides an accurate 
description of Hubble residuals in the Hubble flow. Although the extinction properties assumed in the P23 model are yet to be understood or perhaps verified within the framework 
of independent astrophysical constraints on dust properties (see section~\ref{assumptions}), we use the P23 bias corrections in the Hubble flow as an effective model. 
Based on our data analysis presented in section~\ref{consistency}, we conclude that obtaining equally good fits in the calibration sector and the Hubble flow of the SH0ES 
data requires employing stronger colour corrections in the calibration galaxies than in the Hubble flow galaxies. We achieve this by adjusting the extinction model in the calibration sample. 
The apparent difference between colour (extinction) corrections likely can be attributed to a mismatch between host galaxy morphological types in the calibration sample 
(solely late-type galaxies with well observable, luminous Cepheids) and the training data of the P23 model (both late- and early-type galaxies).

For the new dust model, we assume Milky Way-like $R_{\rm B}$ values given by a Gaussian distribution with the mean value $\langle R_{\rm B}\rangle=4.3$ and 
a scatter of $\sigma_{\rm R_{\rm B}}=0.4$ in both high and low stellar-mass host galaxies. Employing the same distribution in both stellar mass 
bins is expected to eliminate the discrepancies demonstrated in Figure~\ref{anoma_ori} which are most likely related to the mass-step in $\langle R_{\rm B}\rangle$ 
assumed in the P23 model. The adopted mean is given by the measurements of $R_{\rm B}$ in the Milky Way \citep{Schlafly2016,Fitzpatrick2007}. Its value 
also matches the extinction curve ($R_{\rm V}\approx R_{\rm B}-1=3.3$) employed by \citet{Riess2022} in the baseline model to determine colour corrections for Cepheids. This means that 
extinction corrections of both Cepheids and supernovae (via the bias model for the latter) will be based on the same extinction curve model, as opposed to the P23 
dust model which assumes $\langle R_{\rm B}\rangle \approx 3.1$ in $M_{\star}>10^{10}M_{\odot}$ host galaxies. 
The adopted scatter $\sigma_{\rm R_{\rm B}}$ is also motivated by the measurements in the Milky Way for which the dispersion of $R_{\rm B}$ 
ranges from $0.18$ \citep{Schlafly2016}, through $0.3$ \citep{Fitzpatrick2007} to $0.6$ \citep{Legnardi2023}. It is respectively 3 and 2 times smaller 
than the corresponding scatter in low- and high stellar-mass bins of the P23 model. Figure~\ref{RB_models} compares the adopted $R_{\rm B}$ 
distribution to the P23 model and observational constraints from the Milky Way.

\begin{figure*}
    \centering
    \begin{subfigure}[l]{0.45\textwidth}
        \centering
        \includegraphics[width=\linewidth]{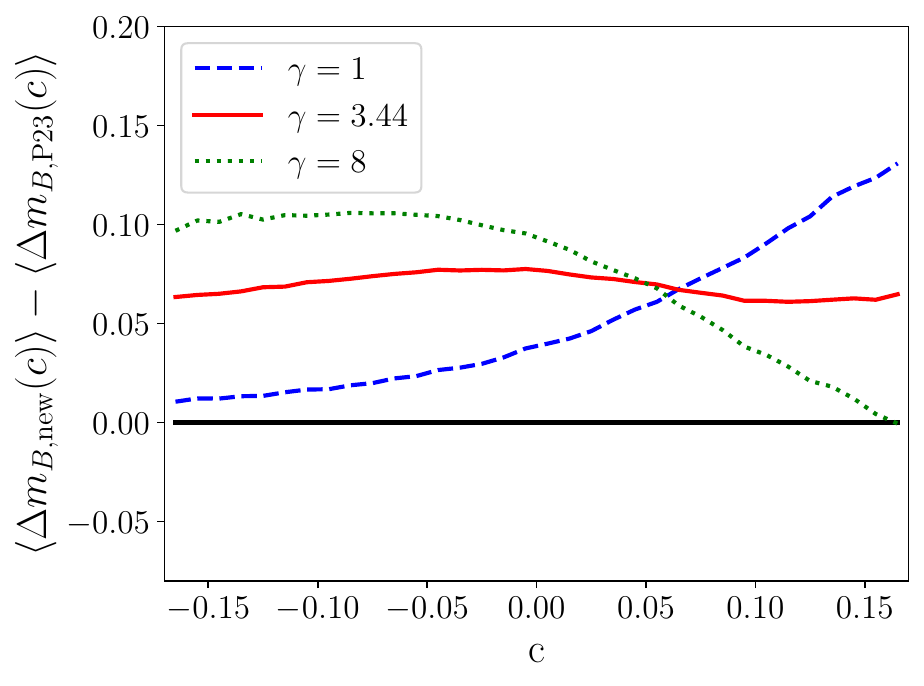} 
        \caption{} \label{gamma_sim_a}
    \end{subfigure}
    \begin{subfigure}[r]{0.44\textwidth}
        \centering
        \includegraphics[width=\linewidth]{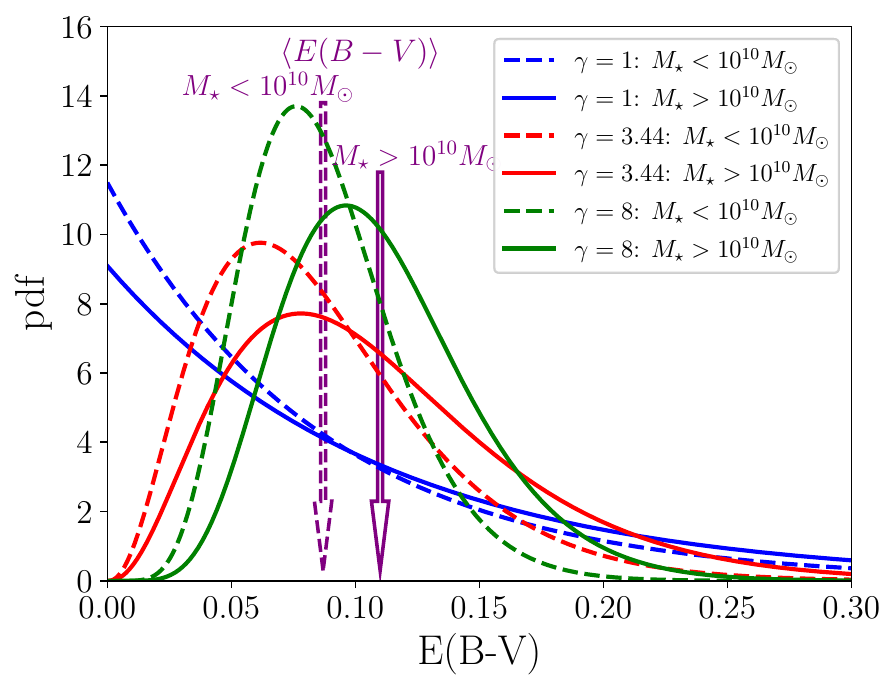} 
        \caption{} \label{gamma_sim_b}
    \end{subfigure}
    \caption{\textit{Left:} Average supernova peak magnitude relative to the P23 model as a function of colour, simulated for the new dust model with different shapes of the reddening distribution given by equation~(\ref{pdf_gamma}). 
    \textit{Right:} The corresponding dust reddening distributions parameterised by the shape parameter $\gamma$. All models preserve the mean dust reddening in the 
    high and low stellar-mass bins of supernova hosts (marked with the purple arrows) from the P23 model \citep{Popovic2021} adopted 
    in the Pantheon+ compilation. The distribution model with $\gamma=3.44$ is found to reproduce closely the colour dependence of the simulated average 
    peak magnitudes in the P23 model and yields the same effective slope $\beta$ of the supernova peak magnitude--colour relation as the P23 model.
    }    \label{gamma_sim}
\end{figure*}

The higher $R_{\rm B}$ values adopted in the new dust model inevitably results in a larger effective slope $\beta$ of the simulated $m_{\rm B}$--$c$ relation. 
We find that $\beta=3.0$ from the P23 model can be easily reproduced by modifying the shape of the assumed distribution of $E(B-V)$ values of the supernovae. A practically 
exact match between the two models in terms of $\beta$ can be achieved using a gamma distribution given by
\begin{equation}
p(E(B-V))=\Big(\frac{E(B-V)}{\tau}\Big)^{\gamma-1}\exp\Big(-\frac{E(B-V)}{\tau}\Big)\frac{1}{\tau},
\label{pdf_gamma}
\end{equation}
which is a generalisation of the exponential model (which is recovered for $\gamma=1$), introduced to hierarchical Bayesian modelling of type Ia supernovae by \citet{Wojtak2023}. 
Varying the shape parameter $\gamma$ and keeping the mean 
reddening $\langle E(B-V)\rangle =\gamma\tau$ fixed at the values of the P23 model, we find that the simulated $\Delta m_{\rm B}$--$c$ relation attains the same 
slope as that of the P23 model for $\gamma=3.44$. From the viewpoint of information encapsulated by model distributions, the proposed modification provides a second order correction (second moment) to the exponential model with a given mean value. It is in this sense that the proposed modification of the reddening distribution is as minimal as possible. From an astrophysical point of view, different values of $\gamma$ reflect different spatial extensions of type Ia supernova positions with respect to a fixed dust disk (with maximum for $\gamma=1$). We discuss the astrophysical implications of finding $\gamma\approx 3.4$ in section~\ref{discussion}.

Figure~\ref{gamma_sim} shows an average simulated peak magnitude as a function of colour parameter $c$ 
(relative to the P23 model) for a range of $\gamma$ values, and the corresponding $E(B-V)$ distributions. The figure demonstrates that the new model with 
$\gamma=3.44$ not only matches the effective slope $\beta$, but it also reproduces the non-linearity of the $\langle \Delta m_{\rm B}\rangle$--$c$ 
relation of the P23 model. In terms of the shape of the predicted $\langle \Delta m_{\rm B}\rangle$--$c$ relation, the models appear to be practically indistinguishable. 
Keeping the slope $\beta$ to be consistent with the P23 model is a conservative assumption required by preserving consistency with the precomputed statistical 
uncertainties $\sigma_{\rm meas}$ (assuming $\beta\approx 3.0$). Tests against the calibration data shown in the following subsections demonstrate that 
this assumption together with the adopted Milky Way-like extinction is favoured by the observations.

\begin{figure*}
    \begin{subfigure}[t]{0.45\textwidth}
        \centering
        \includegraphics[width=\linewidth]{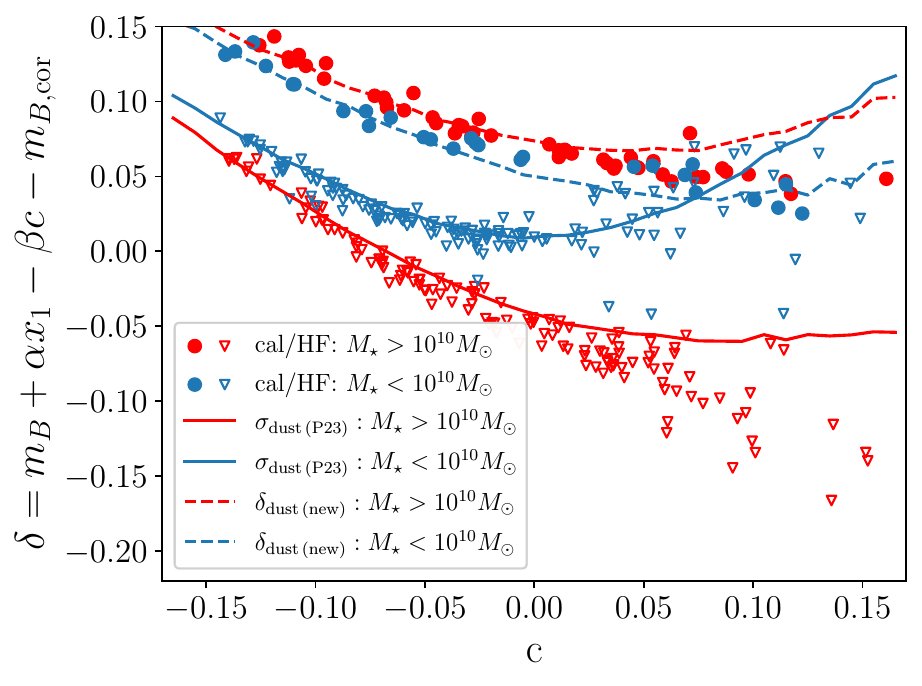} 
        \caption{} \label{HD_ori_b_new}
    \end{subfigure}
    \begin{subfigure}[t]{0.45\textwidth}
    \centering
        \includegraphics[width=0.97\linewidth]{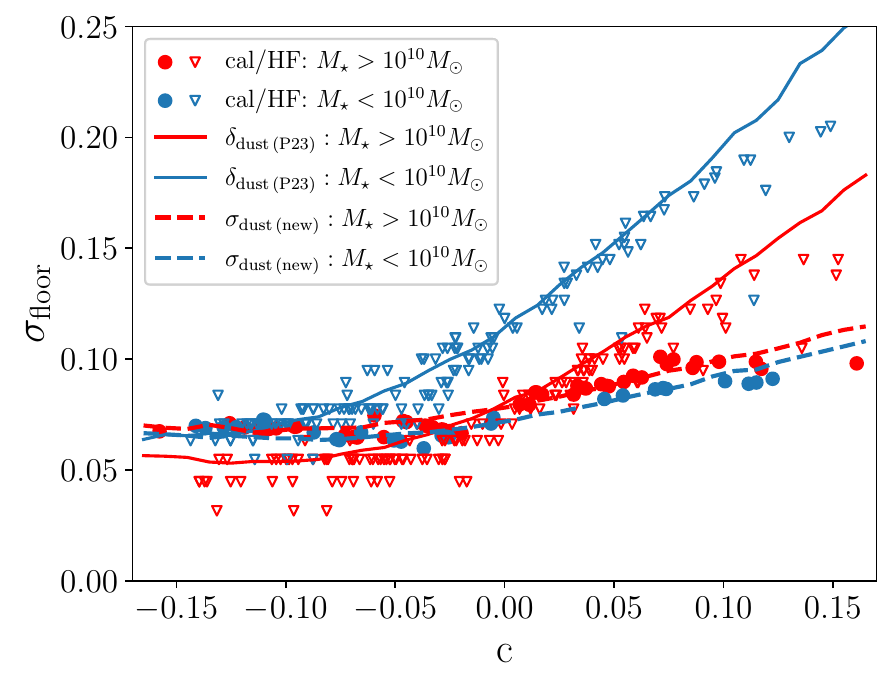} 
        \caption{} \label{HD_ori_c_new}
    \end{subfigure}

    \centering
    \begin{subfigure}[l]{0.6\textwidth}
        \centering
        \includegraphics[width=\linewidth]{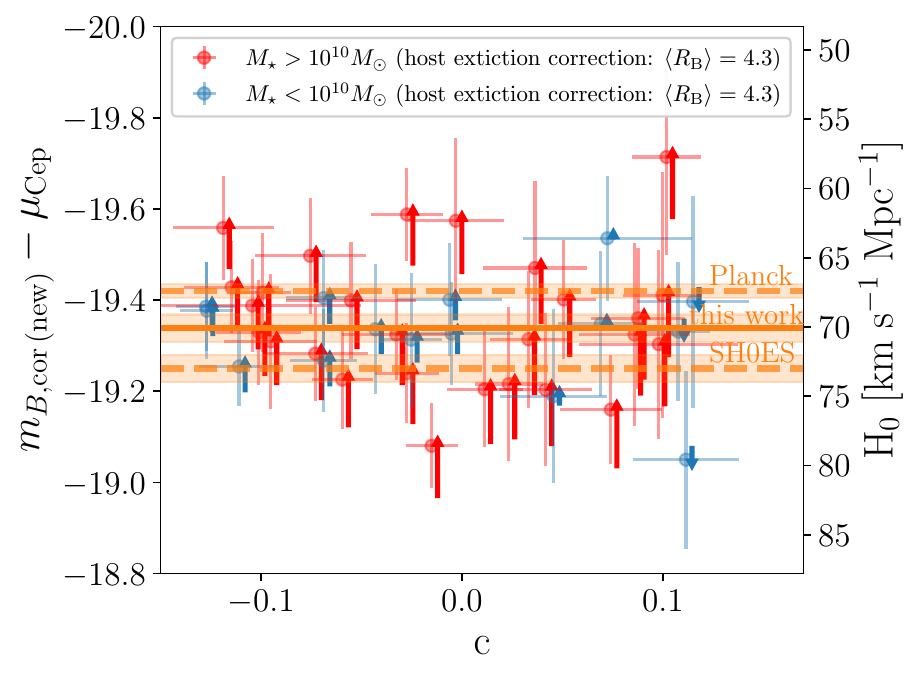} 
        \caption{} \label{HD_ori_a_new}
    \end{subfigure}
    
    \caption{Same as Figure~\ref{HD_ori} but with supernova peak magnitudes corrected using the 
    new dust model developed in section~\ref{new_dust}. The new model applies solely to the calibration sample. The bias corrections 
    and the floor uncertainties in the Hubble flow based on the P23 model remain unchanged. The new model assumes Milky Way-like total-to-selective 
    extinction coefficients with $\langle R_{\rm B}\rangle=4.3$ and $\sigma_{\rm R_{\rm B}}=0.4$ (see Table~\ref{dust-models}), and a modified shape of 
    the $E(B-V)$ distribution (preserving the same mean reddening in both stellar mass bins as in the P23 model and the same slope $\beta$ of the supernova peak magnitude--colour relation; 
    see Figure~\ref{gamma_sim_b}). The dashed and solid curves on the top panels show the bias correction and the floor uncertainties, computed in the new model ($\delta_{\rm dust\,(new)}$, 
    $\sigma_{\rm dust\,(new)}$) and the P23 model ($\delta_{\rm dust\,(P23)}$,$\sigma_{\rm dust\,(P23)}$), respectively. The new model increases the extinction correction, 
    particularly in the high stellar-mass bin, and decreases the floor uncertainties $\sigma_{\rm floor}(c)$ (intrinsic scatter) for red supernovae. The arrows on the bottom panel show the change of supernova 
    corrected magnitudes due to replacing the P23 model with the new dust model. The error bars include corrections due to modified floor uncertainties. The horizontal lines (shaded bands) show the Hubble constant values (uncertainties) from SH0ES 
    \citep[][matching the best fit $M_{\rm B}$ derived from the calibration data]{Riess2022}, Planck \citep{Planck2020_cosmo} and the 
    reanalysis of the SH0ES data presented in this work.}
     \label{HD_new}
\end{figure*}

\subsection{Modified bias and intrinsic scatter}

We compute the bias and the floor uncertainties in the new dust model (see parameters in Table~\ref{dust-models}) using the method based on 
Monte Carlo simulations described in section~\ref{dust_model}. The difference between biases in the new model and the P23 model 
is given by
\begin{eqnarray}
  & \Delta\delta_{\rm dust}=\langle \Delta m_{\rm B\,dust\,(new)}(c,M_{\star})\rangle-\langle \Delta m_{\rm B\,dust\,(P23)}(c,M_{\star})\rangle. \nonumber\\
\end{eqnarray}
The new extinction correction (bias) can be easily incorporated in the original Pantheon+ corrected peak magnitudes $m_{\rm B,corr}$ in the following way
\begin{equation}
m_{\rm B,corr\,(new)}=m_{\rm B,corr}-\Delta\delta_{\rm dust},
\end{equation}
where $m_{\rm B,corr\,(new)}$ is the corrected peak magnitude estimated in the new dust model. The above correction effectively subtracts the original bias from 
the P23 model and applies the one based on the new model. Here we assume that possible survey effects (and their modified estimation given the new model) are negligible. This is a safe assumption given that the original bias estimation from the Pantheon+ catalogue is entirely driven by the extinction effect in the underlying dust model (see Figure~\ref{HD_ori_b}).

Figure~\ref{HD_ori_a_new} compares the new corrected supernova peak magnitudes to distance moduli of Cepheids, while Figure~\ref{HD_ori_b_new} compares 
 the new bias in the calibration sample to the P23 bias, which now applies solely to the Hubble flow supernovae. We note that the biases in both models are calculated with respect to the same value of the reference magnitude $\Delta_{0}$ defined in equation~(\ref{bias_gen}). For the sake of convenience, we choose $\Delta_{0}$ derived from fitting the simulated magnitudes in the P23 model so that $\delta_{\rm dust\,(P23)}$ for supernovae in the Hubble flow remains the same as in Figure~\ref{HD_ori}. 
 The figures show that the bulk modification of bias occurs in the high stellar-mass hosts and it is due to increased mean value 
 of $R_{\rm B}$. The apparent change of $\delta_{\rm dust}$ in the low stellar mass sample is primarily driven by the change of the dust reddening distribution relative to the P23 model. Unsurprisingly, the new model yields nearly the same bias corrections for both stellar mass bins. A mild difference between the profiles reflects a small difference between the mean reddening in the two stellar mass bins, which are chosen to be the same as in the P23 model.

The adopted Milky Way-like scatter in $R_{\rm B}$ values results in significantly lower floor uncertainties, especially for red supernovae with $c>0$ (see Figure~\ref{HD_ori_c_new}). We incorporate the new scatter model in the Pantheon+ supernova covariance matrix by rescaling the estimates of $\sigma_{\rm floor}(c,M_{\star})^{2}$ based on the P23 model 
and provided by the Pantheon+ catalogue ($biasCor\_m\_b\_COVADD$) proportionally to the ratio of $\sigma_{\rm dust}^{2}(c,M_{\star})$ computed in the two dust models. The corrections are implemented using
\begin{eqnarray}
 & \Delta\sigma_{\rm floor}^{2}\equiv \sigma_{\rm floor\,(new)}^{2}(c,M_{\star})-\sigma_{\rm floor\,(P23)}^{2}(c,M_{\star}) \nonumber \\
& \approx \sigma_{\rm floor\,(P23)}^{2}(c,M_{\star})\Big( \frac{ \sigma_{\rm dust\,(new)}^{2}(c,M_{\star})}{ \sigma_{\rm dust\,(P23)}^{2}(c,M_{\star})}-1\Big),\nonumber \\
\end{eqnarray}
which quantifies the difference between the floor uncertainties due to the change of the dust model, given the scatter estimates from the Pantheon+ catalogue 
($\sigma_{\rm floor\,(P23)}$). The modified supernova covariance matrix corresponding to the new dust model is obtained by calculating $\Delta\sigma_{\rm floor}^{2}$ for each supernova and adding this correction to all elements of the covariance matrix containing $\sigma_{\rm floor}^{2}$ (all diagonal elements and off-diagonal elements associated with duplicates of the same supernova). Hereafter, we refer to the resulting covariance matrix as $\mathsf{C}_{\rm stat+syst}^{\rm SN(new)+Cepheid_{\rm cal}}$ in contrast to the original 
Pantheon+ matrix $\mathsf{C}_{\rm stat+syst}^{\rm SN+Cepheid_{\rm cal}}$ in the calibration data sector. We find that the difference between log determinants of the new and original covariance matrices due to decreased floor uncertainties in the supernova sector is
\begin{equation}
\ln\det \mathsf{C}_{\rm stat+syst}^{\rm SN(new)+Cepheid_{\rm cal}}-\ln\det \mathsf{C}_{\rm stat+syst}^{\rm SN+Cepheid_{\rm cal}}\approx-2.6.
\label{delta_det}
\end{equation}

\subsection{Observational tests}\label{observational_tests}

We begin testing the new dust model by verifying that the constraints on $M_{\rm B}$ from supernovae 
$M_{\star}>10^{10}M_{\odot}$ are consistent with those from the entire calibration sample. We proceed with the same 
consistency tests as in section~\ref{tests_ori}, but with the the likelihood replaced by
\begin{eqnarray}
\ln L_{\rm new} & \propto & -\frac{1}{2} \pmb{\Delta\mu_{\rm new}^{T}}(\mathsf{C}_{\rm stat+syst}^{\rm SN(new)+Cepheid_{cal}})^{-1}\pmb{\Delta\mu_{\rm new}}\nonumber\\
&-&\frac{1}{2}\ln\det(\mathsf{C}_{\rm stat+syst}^{\rm SN(new)+Cepheid_{cal}}),\nonumber \\
\end{eqnarray}
where $\Delta\mu_{\rm new,i}=m_{\rm B,corr\,(new),i}-M_{\rm B}-\mu_{\rm Cep,i}$.

As demonstrated in Figure~\ref{anoma_new}, the new model to a large extent alleviates the systematic discrepancies between high and low stellar-mass host galaxies in the original data. The best fit absolute magnitude 
(or its equivalent $H_{0}$ from equation~(\ref{M_B_H_0})) derived from supernovae in high stellar-mass hosts merely differs by $0.7\sigma$ 
(and $1.6\sigma$ when restricting the sample to red supernovae with $c>0$) relative to the mean of the best fit values from random comparison subsamples.

\begin{figure*}
    \centering
    \begin{subfigure}[l]{0.45\textwidth}
        \centering
        \includegraphics[width=\linewidth]{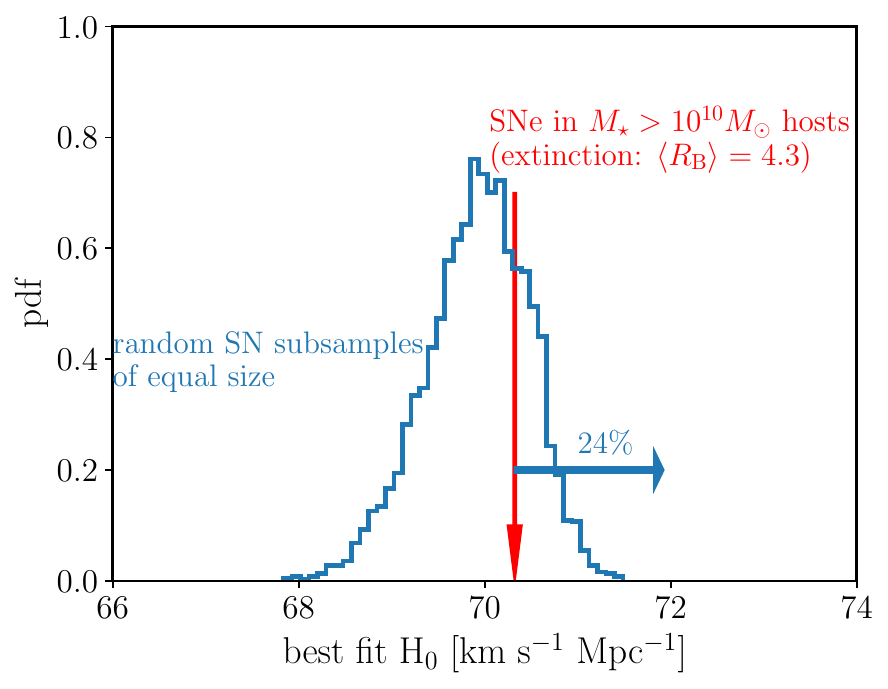} 
        \caption{} \label{anoma_ori_a_new}
    \end{subfigure}
    \begin{subfigure}[r]{0.45\textwidth}
        \centering
        \includegraphics[width=\linewidth]{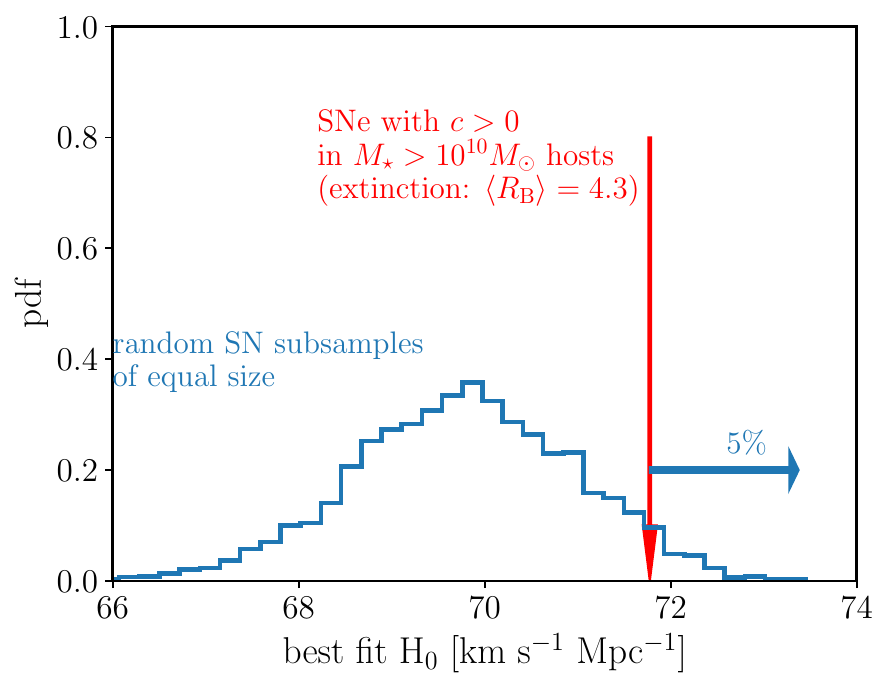} 
        \caption{} \label{anoma_ori_b_new}
    \end{subfigure}\label{anoma_new}
    \caption{Same as Figure~\ref{anoma_ori} but for corrected supernova peak magnitudes and their covariance matrix recomputed in the 
    new dust model which assumes Milky Way-like extinction coefficients $R_{\rm B}$ and a modified shape of the reddening distribution (see Table~\ref{dust-models}). 
    The new model alleviates systematic effects apparent in the data based on the P23 model (see Figure~\ref{anoma_ori}). The best fit $H_{0}$ obtained for supernovae 
    in the high stellar-mass host galaxies is consistent with the control random subsamples at $0.7\sigma$ (and $1.6\sigma$ for its subsample with $c>0$). A substantially 
    lower $H_{0}$ results from a stronger extinction correction in the calibration galaxies, primarily in $M_{\star}>10^{10}M_{\odot}$ host galaxies, 
    than the P23 model.}\label{anoma_new}
\end{figure*}

\begin{table*}
\begin{center}
\begin{tabular}{lcccc}
\hline
dust model (calibration)& \multicolumn{2}{c}{Pantheon+\citep{Popovic2021}} & \multicolumn{2}{c}{this work}  \\
 & & & & \\
 dust model (Hubble flow (HF))& \multicolumn{2}{c}{Pantheon+\citep{Popovic2021}} &  \multicolumn{2}{c}{Pantheon+\citep{Popovic2021}}   \\
 & & & & \\
data set & calibration & calibration+HF & calibration & calibration+HF \\
\hline
$M_{\rm B}$ & $-19.249\pm0.03$ &$-19.248\pm0.03$  & $-19.338\pm0.03$ & $-19.334\pm0.03$\\
$H_{0}$[km\,s$^{-1}$\,Mpc$^{-1}$]  & $73.0\pm1.0^{\rm a}$ & $73.4\pm1.0$ & $70.1\pm1.0^{\rm a}$ & $70.5\pm1.0$\\
$\chi^{2}_{\rm min}$ & 57.1& 300.1 & 48.7 & 285.5 \\
$\Delta$BIC (cal)& 0 & -- & $-11.0$ & --\\
\hline
\end{tabular}
\caption{Results of fitting $M_{\rm B}$ to the calibration data set ('calibration') and $\{M_{\rm B},H_{0}\}$ to the calibration set and the Hubble flow ('calibration+HF'), 
for two different models of dust extinction in the calibration sample (see Table~\ref{dust-models}). The table provides best fit parameters (posterior mean and standard deviations), minimum 
$\chi^{2}$ and the change of the Bayesian Information Criterion $BIC=-2\ln L$ (fixed degrees of freedom) for fits with the calibration data set.\\
$^{a}$ The estimate of the Hubble constant is derived assuming equation~(\ref{M_B_H_0}).
}\label{fits}
\end{center}
\end{table*}

Table~\ref{fits} shows the best fit $M_{\rm B}$ (and equivalent $H_{0}$) measured from the calibration data. We find that the new dust model is 
strongly favoured over the P23 model implemented in the Pantheon+ compilation. The Bayesian Information Criteria (BIC) is 11.0 lower than in the P23 model 
($\Delta BIC=-11.0$), with a contribution of $8.4$ from $\Delta \chi^{2}_{\rm min}$ and $2.6$ from the change in the likelihood normalisation (see equation~\ref{delta_det}). 
The apparent strong preference for the new model is an effect of corrected biases in supernova peak magnitudes and floor uncertainties in the supernova covariance matrix 
(for the latter, via reduced total uncertainties of the corrected peak magnitudes and correlations between duplicates). We emphasise that all 
three new components of our model, i.e. $\langle R_{\rm B}\rangle=4.3$, $\sigma_{\rm R_{\rm B}}=0.4$ and $\gamma=3.44$ (see Table~\ref{dust-models}), 
contribute to the obtained goodness of fit. This can be realised by comparing fits with simplified models retaining one extra component from the P23 model. 
We find that models with $\gamma=1$ or $\sigma_{\rm R_{\rm B}}$ from the Pantheon+ dust model also yield  better fits than the P23 model, 
but with lower significance levels than the complete three-component model: $\Delta BIC=-4.5$ for the former and $\Delta BIC=-3.2$ for the latter. 
As expected, changing solely the shape parameter of the $E(B-V)$ distribution in the P23 model to $\gamma=3.44$ (and retaining the $R_{\rm B}$ distribution) 
increases the discrepancy between absolute magnitudes of blue and red supernovae in high stellar-mass galaxies and thus worsens the fit. Results from the variants of the 
main model described above are summarised in Table~\ref{variants}.

\begin{table}
\begin{center}
\begin{tabular}{ccccc}
\hline
   $R_{\rm B}$ & $\sigma_{\rm R_{\rm B}}$ & $\gamma$ & $\Delta$BIC & $H_{0}$[km\,s$^{-1}$\,Mpc$^{-1}$]$^{\rm a}$ \\
 \hline
   4.3 & 0.4 & 3.44 & $-11.0$ & $70.1\pm1.0^{\rm a}$ \\
   4.3 & 0.4 & (P23) & $-4.5$ & $71.3\pm1.0^{\rm a}$ \\
   4.3 & (P23) & 3.44 & $-3.2$ & $70.0\pm1.0^{\rm a}$ \\
  (P23) & (P23) & 3.44 & +8.4 & $72.8\pm1.0^{\rm a}$\\
  (P23) & (P23) & (P23) & 0 & $73.0\pm1.0^{\rm a}$ \\
\hline
\end{tabular}
\caption{Variants of the main extinction model proposed in this study (included in the first row) and the quality of their fits to the calibration data, as measured by the Bayesian Information Criterion $BIC=-2\ln L$ (for fixed degrees of freedom). The variants retain the values of selected parameters from the P23 model \citep[see Table~\ref{dust-models};][]{Popovic2021}. The main model yields the best fit to the calibration data.\\
$^{a}$ The estimate of the Hubble constant is derived assuming equation~(\ref{M_B_H_0}).
}\label{variants}
\end{center}
\end{table}

\subsection{The Hubble constant}\label{final_H0}
 
Fits using the calibration data only show that the new dust model results in intrinsically brighter type Ia supernovae 
($\Delta M_{\rm B}\approx -0.089$ mag), and hence inevitably a lower value of the Hubble constant ($\delta H_{0}\approx -3$~km\,s$^{-1}$\,Mpc$^{-1}$, 
see Table~\ref{fits}). Here we derive a more rigorous estimate of the Hubble constant based on a joint fit to the calibration data and 
the Hubble flow data comprising 277 supernovae selected by \citet{Riess2022}. The full (77+277)-dimensional covariance matrix 
$\mathsf{C}_{\rm stat+syst}^{\rm SN+Cepheid_{cal+HF}}$ for supernovae both in the calibration sample and the Hubble flow is extracted from 
the Pantheon+ repository (\textit{Pantheon+SH0ES\_STAT+SYS}) and it includes all statistical and systematic uncertainties of Cepheid distance 
moduli and supernova corrected peak magnitudes. The calibration block of the matrix is the same as that used in 
section~\ref{tests_ori}, i.e.\ $\mathsf{C}_{\rm stat+syst}^{\rm SN+Cepheid_{cal}}$. For the new dust model, this block of the matrix 
is modified in the same way as described in section~\ref{new_dust}. The likelihood is given by the following equation
\begin{equation}
\ln L_{\rm cal+HF}\propto -\frac{1}{2} \pmb{\Delta\mu^{T}}(\mathsf{C}_{\rm stat+syst}^{\rm SN+Cepheid_{cal+HF}})^{-1}\pmb{\Delta\mu},
\end{equation}
where $\Delta\mu_{i}=m_{\rm B,corr,i}-M_{\rm B}-\mu_{\rm Cep,i}$ (or $\Delta\mu_{i}=m_{\rm B,corr\,(new),i}-M_{\rm B}-\mu_{\rm Cep,i}$ 
for the new dust model) in the calibration sector and
\begin{eqnarray}
\Delta\mu_{i} & = & m_{\rm B,corr,i}-M_{\rm B}-\mu(z_{\rm i}) \nonumber \\
\mu(z) & = & 5\log_{10}\Big(\frac{c}{H_{0}}z[1+\frac{1}{2}(1-q_{0})z\nonumber \\
 & - & \frac{1}{6}(1-q_{0}-3q_{0}^{2}+j_{0})z^{2}]\Big)+25 \nonumber \\
\end{eqnarray}
in the Hubble flow. We adopt $q_{0}=-0.51$ derived from fitting high-redshift supernovae \citep{Riess2022} and $j_{0}=1$. We emphasise that the new 
dust model applies solely to the calibration sample so that the corrected peak magnitudes from the catalogue remain unchanged for the Hubble flow data.

The results from the joint fits including the calibration sample and the Hubble flow sample are shown in Table~\ref{fits}. The new dust model  
results in lowering the best fit Hubble constant by $2.9$~km\,s$^{-1}$\,Mpc$^{-1}$, from $H_{0}=73.4\pm1.0$~km\,s$^{-1}$\,Mpc$^{-1}$ 
\citep[consistent with the measurement of][]{Brout2022} to $H_{0}=70.5\pm1.0$~km\,s$^{-1}$\,Mpc$^{-1}$. The change of the minimum $\chi^{2}$ 
is slightly larger than for the fits based only on the calibration data.

\section{Discussion}\label{discussion}

\begin{figure}
    \centering
        \includegraphics[width=\linewidth]{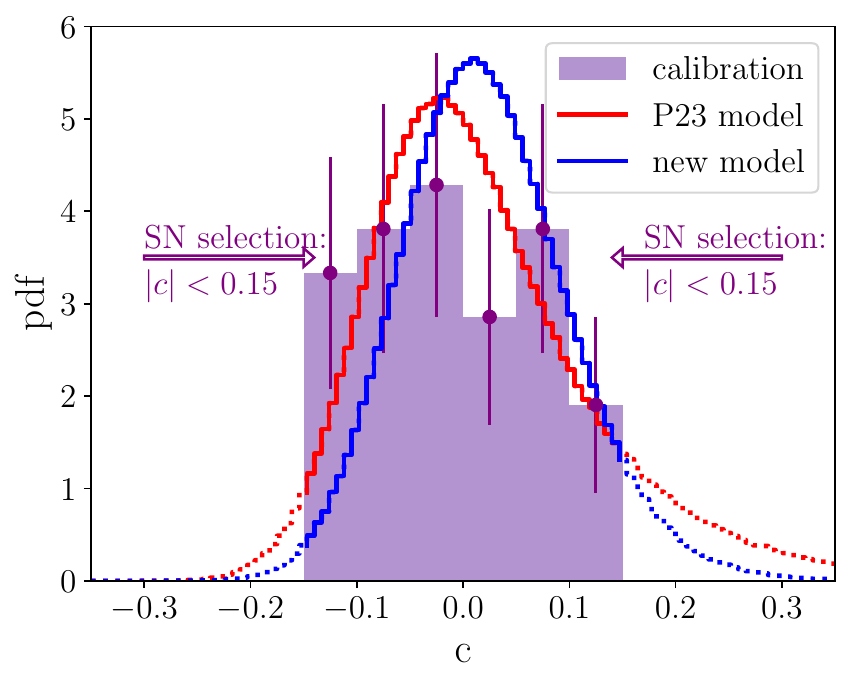} 
    \caption{Distribution of colour parameter $c$ in the calibration sample compared to the colour distributions in the 
    P23 model \citep{Popovic2021} and the new model from this work. The model distributions 
    are normalised to $1$ within the range of observed colours and their tails outside this range are shown as dotted curves. 
    The observed distribution is heavily affected by selection effects including cuts in colours (see the arrows). Testing the two dust models 
    against observations will require using an unbiased supernova sample showing the complete supernova colour 
    distribution in galaxies with observable Cepheids.
    }\label{c-dist}
\end{figure}

The new dust model (Table 1) incorporates Milky Way-like extinction (Figure 3) and a modified shape of the assumed reddening distribution (equation 12). These updates mitigate the 
apparent discrepancies between supernova peak magnitude corrections in the high and low stellar-mass host galaxies, as 
estimated in the Pantheon+ catalogue. A small residual signal can be still seen in red supernovae from the high stellar-mass hosts (see Figure~\ref{anoma_ori_b_new}). 
This may point to a mild underestimation of extinction which can possibly be accounted for in a more refined modelling of the calibration data. 
Ultimately, an optimised model can be obtained by means of full forward modelling of the supernova data in which the
population properties of dust reddening and extinction are constrained independently in the calibration sample and the Hubble flow, 
perhaps using the existing codes for hierarchical Bayesian analysis of type Ia supernovae \citep[see e.g.][]{Mandel2022,Thorp2021,Wojtak2023}.

The minimum $\chi^{2}$ obtained for the fits with the calibration data are $1.5\sigma$ and $2.2\sigma$ lower than the mean 
value expected for $76$ degrees of freedom (77 data points and one free parameter) for the P23 model and the new dust model, respectively. This may indicate a mild overestimation 
of the covariance matrix, e.g.\ through model-dependent elements such as the floor uncertainty $\sigma_{\rm floor}(c)$. Although 
the new dust model reduces $\sigma_{\rm floor}$ and thus the related errors of supernova peak magnitudes, the fit is primarily driven 
by improved residuals in $m_{\rm B,corr\,(new)}-\mu_{\rm Cep}$ and correlations between duplicates, resulting in a lower minimum $\chi^{2}$. The floor uncertainties in the 
new model are primarily dominated by scatter in $\beta_{\rm SN}$ and $c_{\rm int}$. Therefore, further reduction of the floor scatter would require modifications to the prior distributions for supernova intrinsic properties. We also note that unexpectedly low $\chi^{2}$ values also occur for cosmological fits based on the complete 
Pantheon+ data \citep{Keeley2022,Peri2023}. This indicates that the overestimation of the covariance matrix is not limited to the calibration data. \citet{Lovick2023} and \citet{Dainotti2023} proposed to mitigate this problem by rescaling the covariance matrix or employing alternative (non Gaussian) probability 
distribution models for the likelihood.

%Rescaling the covariance matrix or employing alternative 

Although the new dust model implies the same mean colour parameter as in the P23 model, i.e.
\begin{eqnarray}
 & \langle c_{\rm dust(new)}\rangle \equiv \langle c_{\rm int\,dust(new)}\rangle+\langle E(B-V)_{\rm dust(new)}\rangle \nonumber \\
& =\langle c_{\rm dust(P23)}\rangle,\nonumber \\
\end{eqnarray}
the shape of the distribution is different. Figure~\ref{c-dist} shows supernova colour distributions obtained from the two dust models and 
compares to the actual distribution of observed colours in the calibration sample. 
Such a comparison can, in principle, serve as an additional means to test which of the two distributions of reddening is correct. This would require using an unbiased supernova 
sample which would show the complete and unbiased colour distribution in galaxies with observable (not necessarily observed) Cepheids. The current supernova sample 
in the calibration galaxies is, however, heavily affected by selection effects such as cuts in colour parameters with $|c|<0.15$ for the current sample \citep{Riess2022} and 
$|c|<0.10$ for its predecessor with 19 calibration galaxies \citep{Riess2016}. The adopted colour cuts eliminate the tails of the distribution where the 
bulk of the signal differentiating between the dust reddening models occurs.

The Hubble constant measured in our study is fully consistent with the TRGB-based determination of \citet{Freed2019}. It is tempting to think that 
the persistent discrepancy between $H_{0}$ estimates from \citet{Freed2019} and \citet{Riess2022}, confirmed by \citet{Dhawan2023}, may be explained 
as an effect of underestimated extinction in the Cepheid 
calibration galaxies relative to those selected for TRGB observations. The new estimate of the Hubble constant obtained in this work reduces the tension with the Planck $H_{0}$ measurement assuming a flat $\Lambda$CDM cosmology from $5.2\sigma$ to $2.8\sigma$. 
The apparent bias in the $H_{0}$ measurement due to underestimated extinction in the calibration galaxies can be ultimately tested with near infrared observations. 
Although the current $H_{0}$ estimates derived from available near-infrared observations seem to agree with the SH0ES result, the errors are currently too large to be decisive in testing the impact of extinction corrections \citep{Jones2022,Galbany2023}. Due to the scarcity of near-infrared data, the current $H_{0}$ measurements are also strongly dependent on the models employed to measure supernova peak magnitudes or complementary constraints from optical light curves. Unlike supernovae observed in the optical, near-infrared samples are limited to very low redshifts with $z\lesssim 0.04$ \citep{Galbany2023}. For this limited redshift range, the Hubble constant determination relies on modelling peculiar velocity corrections whose current models employed in supernova compilations may be incomplete \citep[see e.g.][]{Sorrenti2024}. Regardless of the above-mentioned limitations of supernova samples from near-infrared observations, we note that both the SH0ES $H_{0}$ measurement and our estimate based on the revised extinction model are within $1\sigma$ range of the measurement obtained by \citet{Galbany2023} based on near-infrared supernova light curves, i.e. $H_{0}=72.3\pm2.0$~km~s$^{-1}$~Mpc$^{-1}$. Applying our extinction correction to the near infrared measurement (about 4 times smaller than in optical) lowers the $H_{0}$ value by $0.7$~km~s$^{-1}$~Mpc$^{-1}$ reducing the difference between our and near-infrared-based $H_{0}$ estimates to $0.5\sigma$.

Our analysis shows that Milky Way-like extinction in the calibration galaxies is strongly favoured over the mass-step model of $R_{\rm B}$ 
in the P23 approach. Unlike the P23 model applied to the high-mass calibration galaxies, the favoured extinction model is fully consistent with the expectations based on 
independent estimates of extinction in star-forming galaxies \citep{Salim2018}. The effective mass-step in 
supernova residuals (averaged over all host galaxies) may be absent in subclasses of morphological types when the stellar mass is not the primary 
variable regulating supernova brightness. An alternative and perhaps more physically motivated choice is a variable measuring the local star formation 
or any related metric \citep{Rigault2013}. Differences between supernovae originating from star-forming and passive environments may occur due to differences in extrinsic properties (dust reddening and extinction) and possibly due to diverse intrinsic properties \citep[supernovae from star-forming/passive environments are linked to single- and double-degenerate progenitor scenarios;][]{Maoz2014}. Observational signatures of 
these differences were recently demonstrated in a two-population hierarchical modelling of type Ia supernova light curve parameters in the Hubble flow 
\citep[][]{Wojtak2023}. In this framework, the mass step can arise as an emergent phenomenon resulting from averaging differences between star-forming and 
passive environments which occur in different proportions in $M_{\star}>10^{10}M_{\odot}$ and $M_{\star}<10^{10}M_{\odot}$ galaxies. Testing this scenario is relevant for understanding both the origin of the mass-step correction and the apparent differences between the extinction model proposed in this work and the P23 model. Modelling physically motivated populations of type Ia supernovae can also alter constraints on intrinsic colours obtained within the P23 framework and their impact on the peak magnitude--colour relation, and the floor uncertainties. For example, slowly declining supernovae, which are expected to dominate in the calibration 
sample due to their occurrence in star-forming environments \citep{Rigault2013}, appear to exhibit bluer intrinsic colours with a scatter that is about twice as small as that in the P23 model \citep{Wojtak2023}.

An exponential model of the $E(B-V)$ distribution employed by \citet{Popovic2021} was  proposed by \citet{Jha2007}, and it is commonly assumed in analyses of type Ia supernova data. The model ascribes the highest probability to sight lines with vanishing dust column density and thus $E(B-V)=0$. It is motivated by simulations assuming a combination 
of extended bulge and thick disk for the spatial distribution of type Ia supernovae in their host galaxies \citep{Hatano1998,Riello2005}. However, our study suggests 
that the calibration supernovae are more consistent with the reddening distribution which peaks at about $E(B-V)\approx0.08$ mag and is well approximated by a gamma 
distribution with $\gamma\approx3.4$. This points to a stronger correlation between dust and supernova spatial distributions in galaxies similar to those in the calibration 
sample than in the simulations motivating the exponential model. Virtually the same distribution of $E(B-V)$ with $\gamma\approx 3$ was also shown to reproduce colours 
of high-stretch (slowly declining light curves) supernovae in the Hubble flow \citep{Wojtak2023}. These supernovae are close analogues of the calibration supernovae 
both in terms of light curve properties and host galaxy types.

\section{Summary and conclusions}

We have tested the extinction model adopted in the Pantheon+ supernova compilation \citep[][P23]{Popovic2021} and used in the recent SH0ES determination 
of the Hubble constant \citep{Riess2022}. We have found that an implicit extrapolation of the P23 model from the training (Hubble flow) data set to the calibration sample gives rise to discrepancies between supernova absolute magnitudes derived from the high stellar-mass ($M_{\star}>10^{10}M_{\odot}$) or low stellar-mass 
($M_{\star}<10^{10}M_{\odot}$) supernova host galaxies. These discrepancies coincide with the different total-to-selective extinction coefficients $R_{\rm B}$ assumed 
by the P23 model for these two stellar mass ranges. We have proposed a new extinction model designed to alleviate this tension. The model entails a 
minimalistic modification of the P23 approach and it can be summarised as follows.
\begin{itemize}
\item It is assumed that the $R_{\rm B}$ values along individual sight lines to supernovae in the calibration galaxies are described by the same Milky Way-like distribution, irrespective of host galaxy stellar mass.

\item The shape (effectively, the second moment) of the dust reddening distribution from the P23 model is modified in a way that that the model preserves 
(i) the effective slope $\beta$ of the supernova peak magnitude-colour relation and (ii) the mean dust reddening of supernovae in the Hubble flow, as given by the P23 model. The obtained $E(B-V)$ distribution resembles closely the distribution inferred from type Ia supernovae in the Hubble flow, with similar high-stretch 
(slowly declining) light curves \citep{Wojtak2023}.

\item The model applies solely to the calibration galaxies. For the Hubble constant determination, type Ia supernovae in the Hubble flow are corrected assuming the P23 approach, used as an effective model which by construction accounts for the biases and intrinsic scatter measured in the Hubble flow, but whose complete physical origin is yet to be understood (see sections \ref{introduction} and \ref{discussion}). The apparent difference between the calibration and Hubble flow galaxies in terms of extinction does not imply different dust properties between the local and more distant universe, but it reflects different selections of supernova host galaxies on the two rungs of the cosmic distance ladder.

\end{itemize}

We have tested the new extinction model using Cepheid and supernova data in the calibration galaxies and obtained a revised determination of the Hubble constant. Our results can be summarised as follows.
\begin{itemize}
\item The new model of extinction correction is strongly favoured by the calibration data (with $\Delta BIC=-11$) over the P23 model. It mitigates 
discrepancies between the low  and high stellar-mass galaxies in terms of residual differences between distance moduli of Cepheids and 
corrected supernova peak magnitudes based on the P23 model. It also brings consistency with typical extinction coefficients $R_{\rm B}$ measured in the Milky 
Way and star-forming galaxies (similar to the calibration sample). 

\item The dispersion of $R_{\rm B}$ values assumed in the new model implies a smaller intrinsic scatter in residual differences between distance moduli of Cepheids and corrected supernova peak magnitudes than the P23 model ($0.08$~mag vs $0.11$~mag, as given by colour averaged floor uncertainties).

\item Supernovae in the calibration galaxies are on average $0.09$~mag intrinsically brighter than in the P23 model. This results in a reduction of the best fit Hubble constant 
by $2.9$~km~s$^{-1}$~Mpc$^{-1}$ and the related Hubble constant tension from $5.2\sigma$ to $2.8\sigma$. The best fit Hubble constant is $70.5\pm1.0$~km~s$^{-1}$~Mpc$^{-1}$, which is 
consistent with the result based on the TRGB observations from \citet{Freed2019}.
\end{itemize}

Our results corroborate and complement the previous findings of \citet{Wojtak2022} based on the earlier version of the Cepheid and supernova data \citep{Riess2016,Riess2019,Riess2021}. 
Our modelling of the supernova peak magnitude--colour relation shows that extinction corrections are sensitive to second-order properties of 
the underlying distributions such as the second moment of the assumed reddening distribution. Extinction-related systematic errors in the $H_{0}$ measurement 
can be eliminated when population properties of dust reddening and the extinction coefficient in the calibration galaxies and the Hubble flow are the same. 
It is conceivable that one can meet this condition by matching supernova host galaxies (or perhaps supernova local environments) from the Hubble 
flow to those from the calibration sample. However, selected analogues of the calibration galaxies should reproduce the entire distribution of the dust reddening and 
the extinction coefficient, at least up to the second moment for the former. This requires using independent observations which can directly constrain both 
dust reddening and extinction coefficients along supernova sight lines.

\section*{Acknowledgments}
This work was supported by research grants (VIL16599,VIL54489) from VILLUM FONDEN. 
RW thanks Albert Sneppen, Asta Heinesen, Darach Watson and Christa Gall for discussions and comments. The authors thank the anonymous referee and Adam Riess for constructive comments that helped improve this work.

\section*{Data availability}
No new data were generated or analysed in support of this research.

\bibliography{master}

\end{document}